\documentclass{llncs}

\usepackage{enumitem}
\usepackage{graphicx} 
\usepackage{amsmath}
\usepackage{booktabs}
\usepackage{amssymb}
\usepackage{subcaption}
\usepackage{tabularx}
\usepackage{float}
\usepackage[hidelinks]{hyperref}
\hypersetup{hypertexnames=false}

\usepackage[utf8]{inputenc}
\usepackage[margin=1.5in]{geometry}

\titlerunning{Estimating Exam Item Difficulty with LLMs}

\title{\texorpdfstring
  {Estimating Exam Item Difficulty with LLMs:\\ A Benchmark on Brazil's ENEM Corpus}
  {Estimating Exam Item Difficulty with LLMs: A Benchmark on Brazil's ENEM Corpus}
}

\author{
Thiago Brant\inst{1,2} \orcidID{0000-0003-4615-1659} \and
Julien K\"uhn\inst{2}\orcidID{0009-0003-6185-6478} \and
Jun Pang\inst{2}\orcidID{0000-0002-4521-4112}
}
\authorrunning{T. Brant et al.}
\institute{
Luxembourg Institute of Socio-Economic Research, Luxembourg \and 
Department of Computer Science, University of Luxembourg, Luxembourg}

\begin{document}
\pagestyle{plain}
\maketitle

\thispagestyle{plain}

\begin{abstract}
As Large Language Models (LLMs) are increasingly deployed to generate educational content, a critical safety question arises: can these models reliably estimate the difficulty of the questions they produce? Using Brazil's high-stakes ENEM exam as a testbed, we benchmark ten proprietary and open-weight LLMs against official Item Response Theory (IRT) parameters for 1,031 questions. We evaluate performance along three axes: absolute calibration, rank fidelity, and context sensitivity across learner backgrounds. Our results reveal a significant trade-off: while the best models achieve moderate rank correlation, they systematically underestimate difficulty and degrade significantly on multimodal items. Crucially, we find that models exhibit limited and inconsistent plasticity when prompted with student demographic cues, suggesting they are not yet ready for context-adaptive personalization. We conclude that LLMs function best as calibrated screeners rather than authoritative oracles, supporting an “evaluation-before-generation” pipeline for responsible assessment design.

\end{abstract}


\section{Introduction}
\label{sec:intro}
Since the rapid rise of large language models (LLMs), much of the excitement in education has focused on their generative capabilities: writing feedback, drafting lesson plans, and increasingly, creating new exam questions on demand \cite{brown2020language,wei2022emergent}. In this emerging ecosystem, it is tempting to plug an LLM into a question-generation pipeline and then rely on a separate, often ad hoc, process to decide which questions are appropriate for students. However, for high-stakes assessment of students, this order of operations is inverted: before we can responsibly generate or select items, we need a reliable way to estimate how difficult a question is, and for whom.

Traditionally, item difficulty is estimated via psychometric models such as Item Response Theory (IRT), using large-scale pilot data from representative samples of students \cite{embretson2000item}. This process is statistically rigorous but operationally expensive: items must be field-tested, calibrated, and periodically refreshed as curricula and cohorts change. If LLMs could approximate item difficulty directly from the question text (and associated material), they would offer a powerful complement to classical pipelines, enabling faster triage of candidate items, earlier quality control for automatically generated questions, and more adaptive assessment design \cite{veeramani2024llm_difficulty_rt}. Yet, treating general-purpose LLMs as psychometrically meaningful difficulty estimators is not straightforward.

Several challenges arise when estimating exam question difficulty with LLMs. First, difficulty is highly domain-dependent: the same model may behave differently for language, humanities, natural science, and mathematics items \cite{hendrycks2020measuring}. Second, many real-world exam questions are multimodal and rely on diagrams, graphs, or tables; compressing these visuals into text risks losing spatial cues that matter for reasoning \cite{subramanian2020limitations}. Third, low prediction error can mask weak rank fidelity: a model might get the average difficulty roughly right while misordering which items are actually easier or harder for students \cite{rogoz-ionescu-2024-unibucllm,zotos2024areyoudoubtful}. Fourth, raw predictions often exhibit global location or scale biases that require calibration before they can be compared to IRT metrics. Finally, small contextual cues in the prompts, for example, mentioning the country or educational background of a student, can shift the output in subtle ways and raise questions of bias and fairness \cite{li2023fairnessllms}.

Most existing work on LLMs in educational assessment has focused on scoring constructed responses, generating feedback, or predicting difficulty on relatively small or synthetic datasets \cite{henkel2023gpt,yaneva2024sharedtask}. Less is known about how contemporary models behave when evaluated against large-scale, high-stakes exam data, and even less about their plasticity: their ability (or inability) to adapt difficulty estimates when given contextual information about different student populations. Yet, this plasticity is crucial for adaptive learning and fair use of AI in education. If LLMs are insensitive to learner context, they risk encoding a “one-size-fits-all” notion of difficulty. If they are overly sensitive or biased, they may amplify existing inequalities when used to drive adaptive testing or content recommendation \cite{baker2022algorithmicbias}.

In this work, we adopt a first-principles perspective on these questions using Brazil's Exame Nacional do Ensino M\'edio (ENEM) as a testbed. ENEM is a nationwide, high-stakes exam taken annually by millions of students, with items calibrated using IRT and spanning four broad subject areas: Languages and Codes (LC), Human Sciences (CH), Natural Sciences (CN), and Mathematics (MT) \cite{inep2009matriz}. We assemble a corpus of 1,031 released ENEM items and evaluate ten contemporary LLMs, both proprietary and open-weight, under eight carefully designed prompt strategies and a closed-loop prompt-evolution variant. Ground truth difficulty is given by official IRT parameters derived from large-scale student performance. This setting allows us to ask not only whether LLMs can approximate item difficulty in absolute terms, but also how well they preserve the relative ordering of items within and across subjects.

Beyond this global view, we take a further step that is particularly important for downstream generative use: we probe the plasticity of these models across different student backgrounds. By conditioning prompts on simple contextual cues that evoke distinct learner populations, for example, students from countries with different achievement profiles or educational systems, we ask whether LLMs meaningfully adjust their difficulty estimates in ways that reflect real performance differences, or whether such cues induce only noisy, model-specific shifts. This analysis is a first step toward understanding whether LLM-based difficulty estimators can support fairness-aware, context-sensitive exam design rather than reinforcing a single, implicit reference group \cite{baker2022algorithmicbias}.

\smallskip
Our contributions in this work are threefold:
\begin{itemize}
  \item \textit{Comprehensive benchmark on a real exam corpus.} We systematically evaluate ten proprietary and open-weight LLMs on predicting IRT-grounded difficulty for 1,031 ENEM items across four subjects, comparing absolute error, rank fidelity, and systematic biases under a common experimental protocol.
  \item \textit{Prompt design, multimodality, and calibration.} We study eight prompt templates and a prompt-evolution variant, quantify the penalty of transcribing visual content into text, and assess how simple post hoc calibration layers can reduce global biases without harming rank correlations.
  \item \textit{Plasticity across student groups.} We introduce a lightweight protocol to probe how difficulty estimates shift when prompts include background cues about different student populations, providing initial evidence on the extent to which current LLMs adapt, or fail to adapt, to learner context in a way that could be useful for fair, adaptive assessment.
\end{itemize}

Taken together, these elements form the backbone of our storyline: before deploying LLMs to generate exam questions at scale, we must first understand their capacity to evaluate question difficulty and to do so in a way that respects the diversity of real students. The technical analyses that follow are intended to serve as building blocks for such reliable LLM-enabled assessment pipelines.

To make our goals explicit, we structure this report around the following research questions:
\begin{itemize}
  \item {\it RQ1: Global difficulty estimation.} To what extent can contemporary LLMs recover IRT-based item difficulty for ENEM questions across subjects, both in absolute terms and in terms of rank ordering?
  \item {\it RQ2: Role of prompts and modality.} How sensitive are LLM difficulty estimates to prompt design and to the presence or absence of visual information (transcribed diagrams, tables, and figures)?
  \item {\it RQ3: Plasticity across student groups.} When we condition prompts on different student backgrounds, do LLMs adjust their difficulty estimates in ways that reflect real performance gaps, or are these shifts limited and noisy?
\end{itemize}

\smallskip
\noindent{\bf Structure of the paper.} Section 2 reviews prior work on LLM-based assessment, psychometric difficulty modeling, multimodality, and demographic or nationality bias. Section 3 describes the ENEM dataset, preprocessing (including visual-to-text transcription), and the IRT-derived 1–10 difficulty targets. Section 4 details our benchmarking protocol: models, prompt families, decoding and parsing, metrics for absolute calibration and rank fidelity, and our post-hoc calibration procedures. Section 5 reports results on overall difficulty estimation and prompt effects (RQ1 \& RQ2), quantifies the multimodal transcription penalty (RQ2), and evaluates calibration improvements (RQ1). Section 6 then probes context sensitivity / plasticity across learner backgrounds using country-cue prompts and fairness diagnostics, addressing RQ3. Section 7 concludes with deployment guidance for an “evaluation-before-generation” pipeline.

\section{Related Work}
\label{sec:related_work}
\subsection{LLMs in Educational Assessment}
\label{sec:llms_edu}
Recent work applies LLMs to assessment tasks that range from scoring and feedback to estimating latent psychometric properties such as item difficulty and response time. Three lines are most relevant here.

\begin{itemize}
\item \textit{End to end prompting.}
Veeramani et al.\ built a two stage pipeline that feeds information extracted by classical NLP tools into an LLM prompt to predict both item difficulty and response time; on the BEA 2024 shared task their best run with Llama 2 reached strong accuracy and generalized to oral narrative items without retraining \cite{veeramani2024bea}.

\item \textit{Hybrid feature plus regressor.}
Razavi and Powers compared direct difficulty ratings from \textit{GPT 4o} with models that first extract rich semantic and cognitive features using the LLM and then fit tree based regressors; this hybrid approach improves correlation with IRT difficulty and lowers error \cite{razavi2025}.

\item \textit{Validation against IRT.}
Domain specific studies benchmark LLM difficulty estimates against IRT ground truth. Jain et al.\ reported that \textit{GPT 4o} and {\it o1} align with two parameter IRT values on reading items but compress extremes, highlighting the gap between answer accuracy and calibrated difficulty \cite{jain2025}. Complementary work treats LLM themselves as respondents: Liu et al.\ fitted IRT to LLM answer logs and showed that ensembles of diverse models better reproduce human ability spreads and produce item parameters that correlate highly with human calibrated values \cite{Liu2025LLMRespondents}. Park et al.\ treated zero shot LLM variants as a cohort of virtual students, fitting IRT curves to their answers to infer difficulty without training data \cite{park2024llmsa}.
\end{itemize}

\paragraph{Implications.}
These studies show that LLMs can provide useful signals for difficulty estimation, that hybridization with classical learners often helps, and that averaging over multiple models can counter model specific calibration bias. Our work complements this by holding to single model predictions under controlled prompts and then applying light calibration, allowing us to isolate how much careful prompting plus post hoc calibration can achieve without ensembling.

\subsection{Psychometric Models and Difficulty Estimation}
\label{sec:psychometric-difficulty}
In educational measurement, classical test theory (CTT) measures difficulty by the item \(p\) value, which is sample dependent and assumes equal discrimination for all items \cite{Hambleton1993,Embretson2000}. Item response theory (IRT) addresses these limits by modeling the probability of a correct response as a function of latent ability and item parameters \cite{Embretson2000}. The 1PL, 2PL, and 3PL models parameterize items with difficulty \(b_i\), discrimination \(a_i\), and guessing \(c_i\) \cite{Rasch1960,Birnbaum1968,BakerKim2004}. Under good model fit, \(b_i\) is effectively sample invariant, which supports large scale testing, adaptive delivery, and equating \cite{LordNovick1968,HambletonSwaminathan1985,DeAyala2009}.

Predicting difficulty before pilot administrations has a long history. Early feature based regressions linked readability, lexical frequency, and syntactic complexity to CTT and IRT difficulty with modest explained variance \cite{Freedle1993,Ferrara2022}. Machine learning approaches use higher dimensional text features and non linear models to improve accuracy across reading and science items \cite{Hsu2018,Stepanek2023,AlKhuzaey2024}. More recently, transformer based methods fine tune encoders to infer IRT parameters or compare zero shot LLMs with smaller domain adapted models; fine tuning often outperforms zero shot prompting for absolute error, while chain of thought adds limited benefit in some settings \cite{Benedetto2020,Benedetto2021,Li2025}.

\paragraph{Implications.}
Text captures meaningful but incomplete information about difficulty. IRT provides a principled target, and both classical features and transformer embeddings are predictive, yet substantial variance remains. This motivates our focus on practical gains from prompt design and simple calibration when fine tuning or large cohorts of human responses are not available.

\subsection{Prompt Engineering and LLM Evaluation}
\label{sec:prompt-engineering}
Prompt design is central to eliciting reliable reasoning. Zero shot and few shot prompts serve as strong baselines. Chain of thought encourages models to show intermediate steps and can improve reasoning performance; a simple trigger phrase can unlock such behavior in zero shot settings \cite{chainofthought,kojima2022large}. Plan and Solve decomposes the task into a plan and execution and reduces logical gaps; self consistency samples multiple reasoning traces and aggregates them to stabilize outputs \cite{wang2023plan,wang2023self}. Persona instructions can change style but do not consistently improve factual performance \cite{zheng2023personas}. Tree of Thought expands chain of thought into a search over intermediate states \cite{tree}.

Closed loop refinement has emerged as a general strategy. AlphaEvolve uses mutation, critique, and selection to evolve solutions beyond strong prompting baselines \cite{novikov2025alphaevolve}. Absolute Zero shows that a single model can synthesize tasks, solve them, and obtain verifiable rewards through code execution, achieving state of the art math and coding accuracy without curated data \cite{zhao2025absolutezero}. Scaling and alignment improve stability but leave blind spots; some studies find areas of low difficulty where larger models remain unreliable \cite{Zhou2024LessReliable}. Errors can also be correlated across models, creating a monoculture risk that limits the benefit of naive ensembling \cite{Kim2025CorrelatedErrors}. For visual inputs, recent evaluations place \textit{GPT 4o} and Gemini Flash among the strongest captioners while also noting that integrating narrative context with diagrams can hurt answer accuracy, underscoring the difficulty of fusing modalities in educational content \cite{ramachandran2025vlm,alawwad2025tqa}.

\paragraph{Implications.}
We operationalize these insights by systematically comparing families of concise prompts, employing a point-based planning strategy prior to response generation when appropriate, and sampling multiple reasoning traces to stabilize model outputs. Subsequently, we apply a lightweight calibration layer. To minimize interference, we maintain a strict separation between image-to-text conversion and difficulty-specific prompting.

\subsection{Geographic, Cultural, and Nationality Bias in LLMs}
\label{sec:geo_cultural_bias}
LLMs can exhibit systematic differences in outputs conditioned on geography, culture, or nationality tokens even when the task and the wording remain constant. Surveys emphasize group conditioned performance gaps as key fairness constructs and recommend isolating demographic tokens during evaluation \cite{Gallegos2024BiasSurvey}. Empirical studies report geographic and nationality biases and show that simple location cues can induce consistent shifts \cite{Manvi2024GeoBias,Zhu2024NationalityLREC}. Cultural alignment can vary with dominant language and pretraining mix, and cultural prompting does not uniformly improve alignment across countries \cite{Tao2024PNASNexus,Alkhamissi2024ACL,Zhao2024WorldValuesBench}.

\paragraph{Implications.}
Following these guidelines, the country-cue probe introduces a single token while keeping all other prompt tokens fixed. We observe small yet systematic shifts that are dependent on both model and subject, consistent with prior findings. These results motivate the inclusion of fairness diagnostics alongside aggregate accuracy metrics.

\section{Dataset and Data Processing}
\label{sec:dataset}
We use items from Brazil's national high school exam (ENEM), years 2017–2022 for both exam days. After removing the five foreign language items per year from the `Languages' section, the final set contains \text{1{,}031} multiple choice questions across four subject areas: Mathematics (MT), Natural Sciences (CN), Human Sciences (CH), and Languages and Codes (LC). Items often include figures, tables, graphs, or photographs, especially in CN and MT.

\smallskip
\noindent\textbf{Extraction and normalization.}
When PDFs exposed a reliable text layer, we extracted text directly; otherwise, we took high resolution crops and used vision capable LLMs to transcribe content to structured text. We prompted \textit{GPT\mbox{-}4o} and \textit{Gemini-2.0-Flash} to reproduce layout, equations (\LaTeX-style when helpful), and all answer options. We selected the more faithful transcription by manual spot checks, with \textit{GPT\mbox{-}4o} chosen for the majority of items due to more consistent formatting. Typical residual slips included object or symbol misidentification, numeric value slips, and missing local visual cues. We intentionally left minor inaccuracies in place to reflect realistic deployment. All items were translated to English with DeepL to standardize inputs across models. Each item was stored as JSON with year, day, subject, question text, presence of visual material, and the correct option.

\smallskip
\noindent\textbf{Ground truth.}
We evaluate against official ENEM 3PL IRT metadata released by INEP. For item $i$ with parameters $(a_i,b_i,c_i)$ and ability $\theta$, the probability of a correct response is
\[
P(\theta)=c_i + \frac{1-c_i}{1+\exp\{-a_i(\theta-b_i)\}},
\]
and we use the difficulty parameter $b_i$ as ground truth. For comparability with model outputs we map $b_i$ to a 1–10 scale via a linear transform (clip to $[1,10]$):
\[
D^{\text{IRT}}_i=\mathrm{clip}\!\left(1 + 9\cdot\frac{b_i+3}{6},\,1,\,10\right).
\]

\smallskip
\noindent\textbf{Dataset summary.}
Table~\ref{tab:dataset} reports per subject counts and the share of items tagged with visual content. Visual prevalence is highest in MT and CN, which we later use to analyze modality effects.

\begin{table}[!t]
\centering
\caption{Dataset statistics by subject. `Visual' indicates items tagged with diagrams, figures, tables, or photographs used as part of the stem or options. 
}
\label{tab:dataset}
\begin{tabular}{lrrrr}
\toprule
\textbf{Subject} & \textbf{\# Items} & \textbf{Visual} & \textbf{Non\textendash visual} & \textbf{\% Visual} \\
\midrule
Mathematics (MT)      & 264 & 165 &  99 & 62.5\% \\
Natural Sciences (CN) & 263 & 134 & 129 & 51.0\% \\
Human Sciences (CH)   & 267 &  34 & 233 & 12.7\% \\
Languages and Codes (LC) & 237 &  60 & 177 & 25.3\% \\
\midrule
\textbf{Total}        & 1{,}031 & 393 & 638 & 38.1\% \\
\bottomrule
\end{tabular}
\end{table}

The combination of translation, image-to-text transcription, and minor residual errors creates a realistic text-only pipeline for models that do not accept images. This enables a clean comparison between items that originally contained visuals and those that did not, which we will leverage in Section~\ref{sec:results_modality} to quantify the modality gap. Using IRT $b$ values aligns estimates to a psychometrically meaningful scale and supports rank-based evaluation in addition to error metrics.

\section{Methods}
\label{sec:methods}

We estimate ENEM item difficulty with general-purpose language models under controlled prompting. This study does not generate new items: all evaluations are conducted on released ENEM questions with IRT-derived ground truth. We test eight prompt families ranging from minimal instruction to structured reasoning and query ten heterogeneous models (closed and open-weight) through a unified pipeline. Model outputs are mapped to a common 1–10 scale and compared against IRT difficulty using absolute error metrics and rank-fidelity measures. Finally, we examine light post hoc calibration and a small country-cue probe.

\subsection{Prompt Families}
\label{sec:prompts}
We compare eight prompt families that span key design dimensions: context provision, role conditioning, explicit reasoning scaffolds, and machine-readable output.

\begin{itemize}
\item \textbf{Zero shot}: minimal instruction, integer output in \([1,10]\).
\item \textbf{Enhanced zero shot}: adding one paragraph defining the ENEM task and what counts as easy vs hard.
\item \textbf{Few shot}: two labeled ENEM exemplars before the target item.
\item \textbf{Persona+few shot}: same as few shot but with an education specialist role instruction.
\item \textbf{Chain of thought}: model reflects on salient features then gives a single integer.
\item \textbf{Point based}: bullet plan over criteria (reading complexity, background knowledge, reasoning steps, distractor quality) then one integer.
\item \textbf{Tree of thoughts}: three short independent rationales, then a reconciled integer.
\item \textbf{Synthetic/Playground}: structured instructions with a fixed JSON key for the integer.
\end{itemize}

Appendix \ref{appendix:prompts} provides a single, consolidated listing of all canonical templates.

\subsection{Model Roster}
\label{sec:models}
We evaluate four proprietary LLM APIs and six open weight models that together span size, training style, and access constraints which are listed in Table~\ref{tab:model-roster}. The exact endpoints and licenses are listed for reproducibility.

\begin{table}[H]
\centering
\caption{Context windows correspond to those used or advertised for our runs.}
\label{tab:model-roster}
\begin{tabular}{l l l r}
\toprule
\textbf{Name} & \textbf{Provider} & \textbf{Type} & \textbf{Context} \\
\midrule
GPT-4o (\texttt{gpt-4o-2024-08-06}) & OpenAI & proprietary & 128k \\
GPT-o3 (\texttt{o3-2025-04-16}) & OpenAI & proprietary & n/a  \\
Claude-4-Sonnet (\texttt{claude-sonnet-4-20250514}) & Anthropic & proprietary & 200k \\
Gemini-2.0-Flash (\texttt{gemini-2.0-flash}) & Google & proprietary &1M \\
Gemini-2.5-Flash (\texttt{gemini-2.5-flash-preview-04-17}) & Google & proprietary & 1M \\
Phi-4 (14B) & Microsoft & open weight & 16k \\
Mistral Large 2411 & Mistral & open weight & 128k \\
DeepSeek-R1 & DeepSeek & open weight & n/a  \\
Qwen-3 14B & Alibaba & open weight & 32k  \\
Llama-3.2 3B Instruct & Meta & open weight & 8k+  \\
\bottomrule
\end{tabular}
\end{table}

\subsection{Prediction Pipeline}
\label{sec:pipeline}
All runs use a unified Python pipeline on ULHPC at the University of Luxembourg~\cite{hpc}. Each prompt is programmatically populated from JSON fields (question text, options, and image or table descriptions when present) and sent to the target model. We isolate completions: batch size 1, no cross-item context, and immediate CSV logging per item to avoid loss on preemption. For items with visuals, captions produced during the data processing steps are concatenated as plain text.

\subsection{Decoding Settings}
\label{sec:decoding}
We favor stable classification-style decoding. API models use temperature \(0.2\) when configurable; OpenAI \textit{o3} uses its fixed setting. Local open weights mirror low-temperature defaults; Qwen 3 14B uses \(\texttt{temperature}=0.6\), \(\texttt{top\_p}=0.95\), \(\texttt{top\_k}=20\) to avoid known degeneration at \(0\). Token limits were generous to prevent truncation for reasoning prompts. 

\subsection{Output Parsing}
\label{sec:postproc}
Models are instructed to output a single integer. We parse the first valid integer in \([1,10]\) with prompt-specific regex patterns. Raw text is preserved alongside the parsed score for audit. Rows without a valid integer after manual spot checks are marked \texttt{NA} and excluded from metric denominators.

\subsection{Metrics}
\label{sec:metrics}
We report Root Mean Squared Error (RMSE) and Spearman rank correlation \(\rho\) between predictions \(\hat{y}\) and IRT-based targets \(y\) on the common 1–10 scale:
\[
\mathrm{RMSE} = \sqrt{\frac{1}{N}\sum_{i=1}^N (y_i-\hat{y}_i)^2},\qquad
\rho = 1 - \frac{6\sum d_i^2}{N(N^2-1)}.
\]
MAE, exact match (EM), and within 1 accuracy (W1A) are reported as descriptive summaries. Significance uses Welch \(t\) tests for pairwise and one-way ANOVA with post hoc comparisons for multi-model or multi-prompt groups; exact tests are reported with results.
\paragraph{Signed residuals and deltas.}
Let $y_i \in [1,10]$ denote the IRT-grounded difficulty label for item $i$ and $\hat{y}_i$ the model prediction on the same scale.
We define the signed residual as
\[
r_i := \hat{y}_i - y_i .
\]
Thus, $r_i>0$ means the model predicts the item as \emph{harder} than IRT (overestimation), while $r_i<0$ means it predicts the item as \emph{easier} than IRT (underestimation).
For any condition contrast $A$ vs.\ $B$ (e.g., visual vs.\ non-visual, or with-cue vs.\ no-cue), we report deltas as differences in means, $\Delta := \mathbb{E}[r\mid A]-\mathbb{E}[r\mid B]$, so $\Delta>0$ indicates a shift toward predicting higher difficulty under condition $A$.

To prevent inflation of false positives due to the large number of slices evaluated, we pre-specify a single primary endpoint and contrast for inferential reporting. The primary endpoint is Spearman rank correlation ($\rho$) between predicted difficulty and IRT-derived difficulty on the full set of 1,031 items. The primary contrast compares models under a single, pre-declared prompt family (Point-based), evaluating whether models differ in rank fidelity on a common protocol. All other endpoints (RMSE/MAE/EM/W1A), prompt families, and subgroup analyses are treated as secondary or exploratory.
Because predictions are paired at the item level (the same items are evaluated across conditions), inferential comparisons use paired resampling or paired tests. For the primary endpoint (difference in Spearman $\rho$), we use an item-level permutation test (paired by item). For secondary error metrics (RMSE/MAE), we compare paired per-item errors using Wilcoxon signed-rank tests. We control family-wise error for the primary family of tests using Holm–Bonferroni. For secondary/exploratory result families (e.g., subject-specific heatmaps, modality splits, and prompt grids), we report BH-adjusted q-values within each family and interpret them as exploratory.

\subsection{Post Hoc Calibration}
\label{sec:calibration}
Raw predictions show a uniform downward shift relative to IRT (see details in Section~\ref{sec:results}). Thus, we fit a minimal shift
\[
y^{\ast}=\hat{y}+c,
\]
with \(c\) estimated on a calibration split to minimize squared error, and report metrics on held-out items via 5-fold cross-validation stratified by year and subject. For completeness, we also fit an affine map
\[
y^{\ast}=a\,\hat{y}+c,
\]
which further reduces RMSE for most models; ranks are unchanged by a pure shift and empirically stable under affine.

\subsection{Bias Probe}
\label{sec:bias}
To test context sensitivity, we inject a single country token into an otherwise fixed instruction and rerun predictions for four representative models (GPT 4o, o3, Phi 4, Llama 3) on all 1031 items across seven countries. For item \(i\), model \(m\), country \(c\), we define the signed residual.
\[
\Delta_{i,m,c}=y_i-\hat{y}_{i,m,c}.
\]
We report mean \(\Delta\) by subject and country, a plasticity index (standard deviation of country means), a quasi linear variance (average within item variance across countries), and a two way ANOVA over \(\Delta\) with fixed effects for model and country. We correlate country means with PISA ranks for descriptive context.

\subsection{Limitations}
\label{sec:limits}
Compute limitations constrained several open-weight runs and necessitated using API access for Mistral Large and DeepSeek R1. Budget constraints similarly reduced the number of prompt families tested with Claude and curtailed the breadth of the bias probe. Our image-to-text captions deliberately preserve residual errors, which we quantify later as a modality effect. These decisions prioritize realism and practical constraints without altering the core claims we evaluate.

\section{Results}
\label{sec:results}

Experiments used the ENEM benchmark of 1,031 multiple-choice questions from Brazil's 2017–2022 national exams, spanning Mathematics, Natural Sciences, Languages, and Human Sciences (270 items each). Ten LLMs were tested under eight prompt strategies (76 model–prompt pairs; \textit{Claude-4-Sonnet} only four due to cost). Models were evaluated in zero/few-shot settings without fine-tuning using RMSE, MAE, Spearman's $\rho$, Exact-Match (EM), and Within-1 Accuracy (W1A) against IRT-based difficulty scores (1–10 scale). Proprietary models ran via API; open-weight ones on the HPC platform at the University of Luxembourg. A targeted bias study (see Section~\ref{sec:bias-analysis}) re-evaluates four representative models with country-token variations across seven nations.

\subsection{Prompt Evolution Experiments}
\label{sec:results-prompt-evolution}

Two evolutionary searches were conducted. A seeded run (initialised with persona\_few\_shot) maintained identical structure and RMSE (1.81) across 30 generations. An unseeded run converged by generation~3 to a similar tutor-persona format with no further improvement.  

Both searches collapsed to variants of the same prompt, indicating that Gemini Flash's instruction tuning already captures the reasoning patterns encoded in few-shot persona prompts. The result suggests diminishing returns for automated prompt optimisation in large models, while smaller or open-weight models may still benefit from explicit reasoning scaffolds.

\subsection{Overall Comparison across Models and Prompts}\label{sec:results_overall}
We start with an overall comparison across all model--prompt configurations to provide a high-level map of the results before turning to more detailed breakdowns.
The goal here is not to over-interpret small differences, but to identify consistent patterns in how models respond to different prompt families and to clarify which choices matter most.
Table~\ref{tab:overall_best} summarizes, for each model, the prompt family associated with its best aggregate performance, thereby separating ``which model'' effects from ``which prompt'' effects.
To avoid collapsing performance into a single number, we report two complementary metrics that capture different aspects of quality: RMSE measures absolute calibration (how close predicted difficulty is to the reference values), while Spearman's $\rho$ measures rank fidelity (how well the configuration preserves the relative ordering of item difficulty).
Considering both metrics side by side makes the trade-offs explicit and motivates the model- and prompt-specific analyses presented in the subsequent sections.

\begin{table}[H]
  \centering
  \caption{Best-performing prompt for each model on the full test set.
           Metrics: \textbf{RMSE} (lower is better) and
           \textbf{Spearman~$\rho$} (higher is better).
           The mean predictor attains the lowest RMSE (1.19) by always predicting the average difficulty, but yields no rank discrimination ($\rho = 0$), illustrating why both metrics are needed.
           Full results appear in Appendix~\ref{app:overall-table}.}
  \label{tab:overall_best}
  \small
  \begin{tabular}{lccc}
     \toprule
     Model & Best prompt & RMSE $\downarrow$ & Spearman~$\rho$ $\uparrow$ \\
     \midrule
     \multicolumn{4}{c}{\emph{Proprietary models}} \\[2pt]
     Claude-4-Sonnet            & enhanced\_zero\_shot & 1.83 & 0.33 \\
     GPT-4o                     & point\_based         & 2.06 & 0.32 \\
     o3                         & point\_based         & 2.25 & 0.31 \\
     Gemini-2.0-Flash           & zero\_shot           & 1.70 & 0.20 \\
     Gemini-2.5-Flash           & point\_based         & 1.97 & \underline{0.37} \\
     Mistral-Large              & enhanced\_zero\_shot & 2.00 & 0.34 \\[3pt]
     \multicolumn{4}{c}{\emph{Open-weight models}} \\[2pt]
     Phi-4                      & persona\_few\_shot   & \textbf{1.47} & \textbf{0.31} \\
     DeepSeek-R1                & enhanced\_zero\_shot & 1.82 & 0.35 \\
     Qwen-3                     & persona\_few\_shot   & 1.91 & 0.11 \\
     Llama-3                    & few\_shot            & \underline{1.40} & 0.10 \\[3pt]
     \multicolumn{4}{c}{\emph{Baseline}} \\[2pt]
     Mean predictor             & --                   & 1.19 & 0.00 \\
     \bottomrule
  \end{tabular}
\end{table}

\clearpage
\begin{figure}[H]
  \centering
  \begin{subfigure}[t]{1\linewidth}
    \centering
    \includegraphics[width=\linewidth]{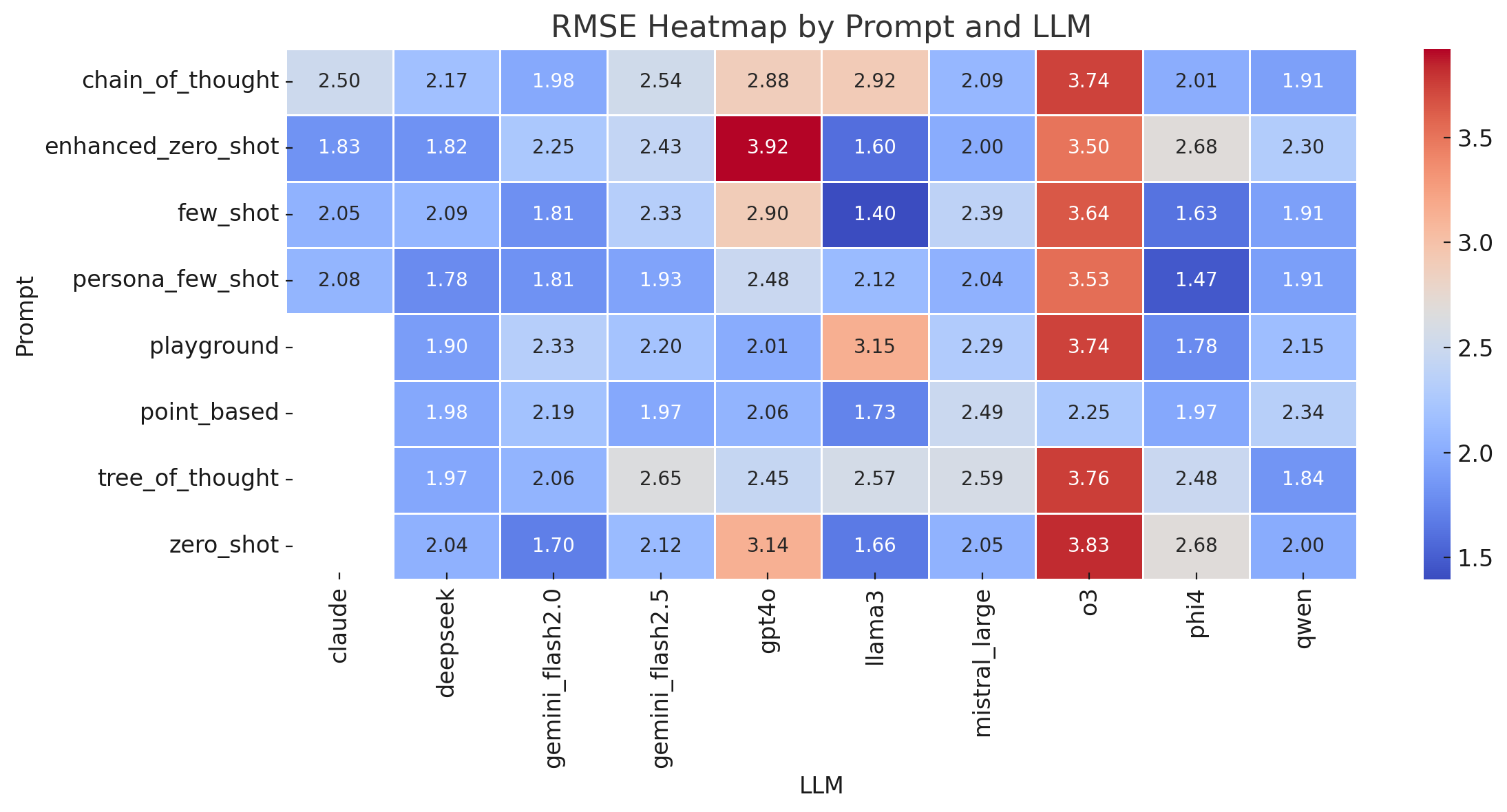}
    \caption{RMSE heatmap (lower is better).}
    \label{fig:overall_heatmap_rmse}
  \end{subfigure}
  \begin{subfigure}[t]{1\linewidth}
    \centering
    \includegraphics[width=\linewidth]{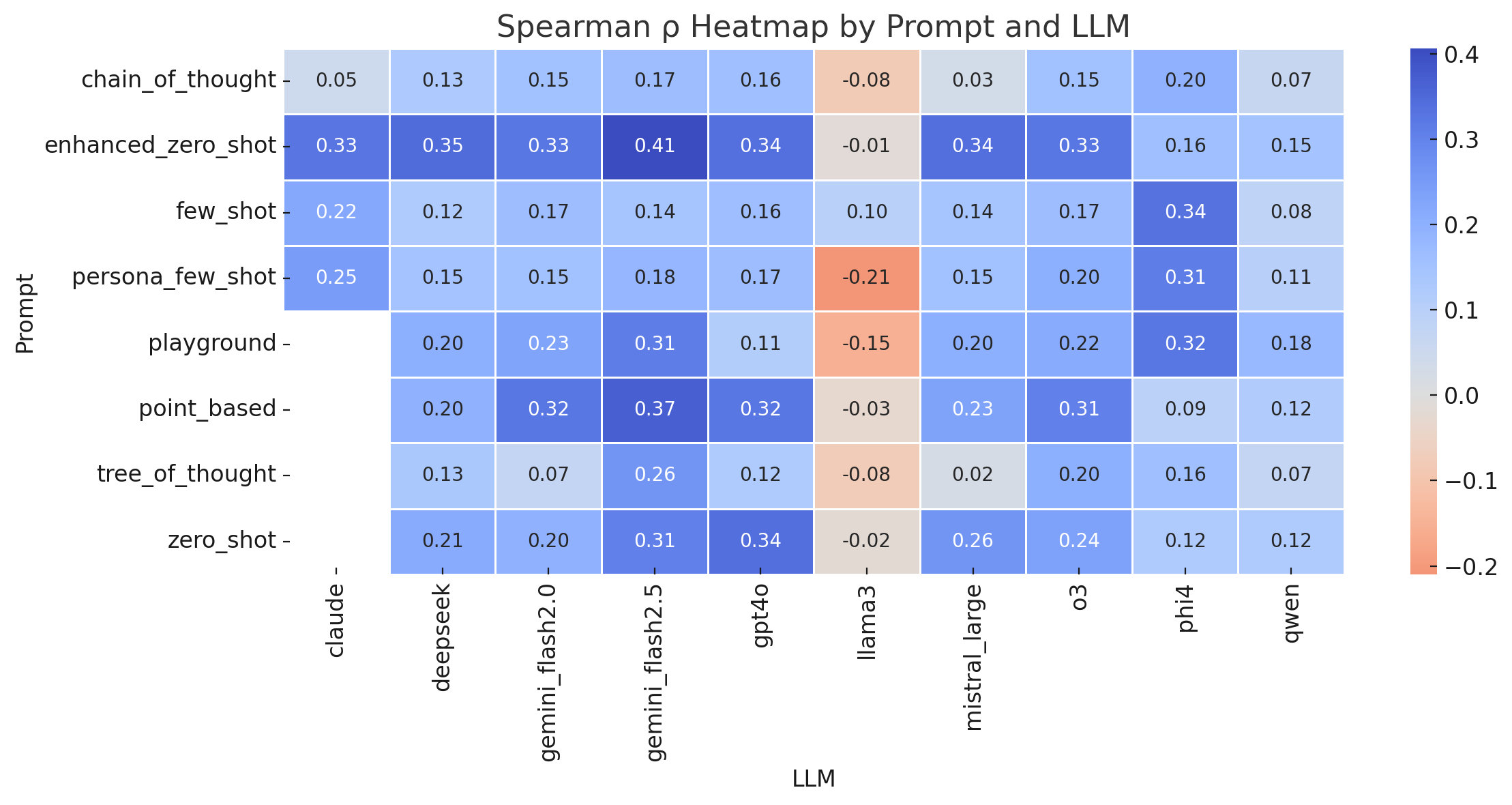}
    \caption{Spearman~$\rho$ heatmap (higher is better).}
    \label{fig:overall_heatmap_spearman}
  \end{subfigure}
  \caption{Performance across all model–prompt pairs.}
  \label{fig:overall_heatmaps}
\end{figure}

\smallskip
\begin{figure}[H]
  \centering
  \includegraphics[width=.9\linewidth]{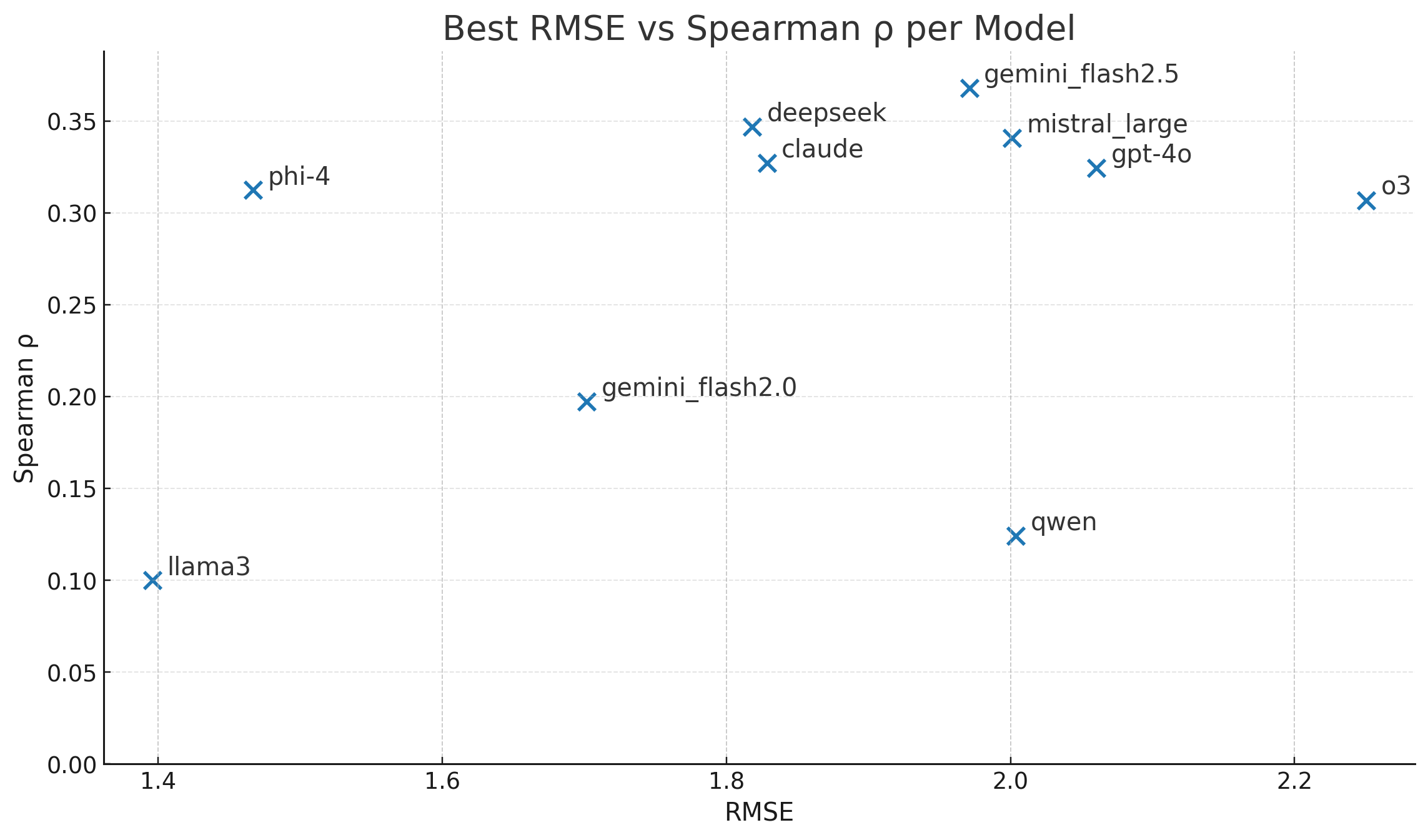}
  \caption{Trade-off between RMSE and Spearman~$\rho$ for each model's best prompt.}
  \label{fig:rmse_vs_rho}
\end{figure}

\vspace{1cm}
The results in Table~\ref{tab:overall_best}  and Fig~\ref{fig:overall_heatmaps}
reveal a clear trade-off between absolute accuracy (RMSE) and rank fidelity (Spearman~$\rho$), and show large within-model prompt sensitivity: for several models, switching prompt families shift performance by amounts comparable to (or larger than) switching models. Although \textit{Llama-3+few\_shot} achieves the lowest RMSE (1.40), its $\rho$ of 0.10 shows that predictions cluster near the mean, limiting discrimination between question difficulties. In contrast, \textit{Phi-4+persona\_few\_shot} slightly increases RMSE (1.47) but improves $\rho$ to 0.31, producing a more realistic ordering of item difficulty (also see Fig.~\ref{fig:rmse_vs_rho}).

Two consistent patterns emerge.
First, \emph{low RMSE can be misleading}.
        Models like \textit{Llama-3} minimise error by regressing toward the 1–10 midpoint, similar to the mean baseline, which results in poor ranking capability and limited pedagogical value.
  Second, \emph{reasoning prompts primarily benefit open-weight models}.  
        For models such as \textit{Phi-4}, \textit{DeepSeek-R1}, and \textit{Qwen-3}, persona- or reasoning-based prompts improve $\rho$ by up to 0.19. In contrast, large instruction-tuned systems (\textit{GPT-4o}, \textit{Llama-3}) show minimal or negative effects, indicating that reasoning scaffolds are most useful where internal reasoning patterns are weaker.

Overall, \textit{Phi-4} provides the most balanced performance, achieving competitive RMSE with strong rank fidelity. These trends are further examined across subject domains (see Section~\ref{sec:results_subjects}) and question modalities (see Section~\ref{sec:results_modality}).

\subsection{Subject Specific Performance}
\label{sec:results_subjects}

We analyse performance by subject (each with \(N{=}270\) items), asking:
(i) which prompt minimises RMSE and maximises Spearman's \(\rho\), and
(ii) which model is most reliable in that domain. Unless noted, numbers are means over all model–prompt pairs; full tables are presetned in Appendix~\ref{app:overall-table}.

\begin{figure}[H]
  \centering
  \includegraphics[width=1.1\linewidth]{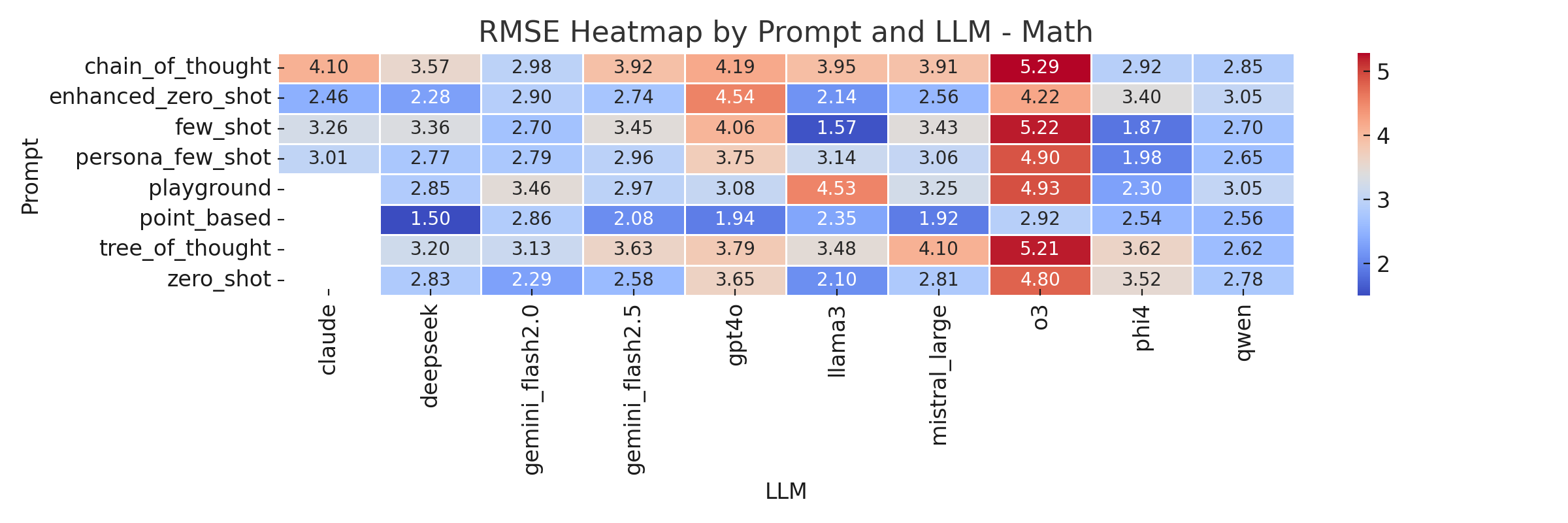}
  \caption{Mathematics: RMSE heatmap by prompt and model.}
  \label{fig:math_heatmap_rmse}
\end{figure}

\begin{figure}[H]
  \centering
  \includegraphics[width=1.1\linewidth]{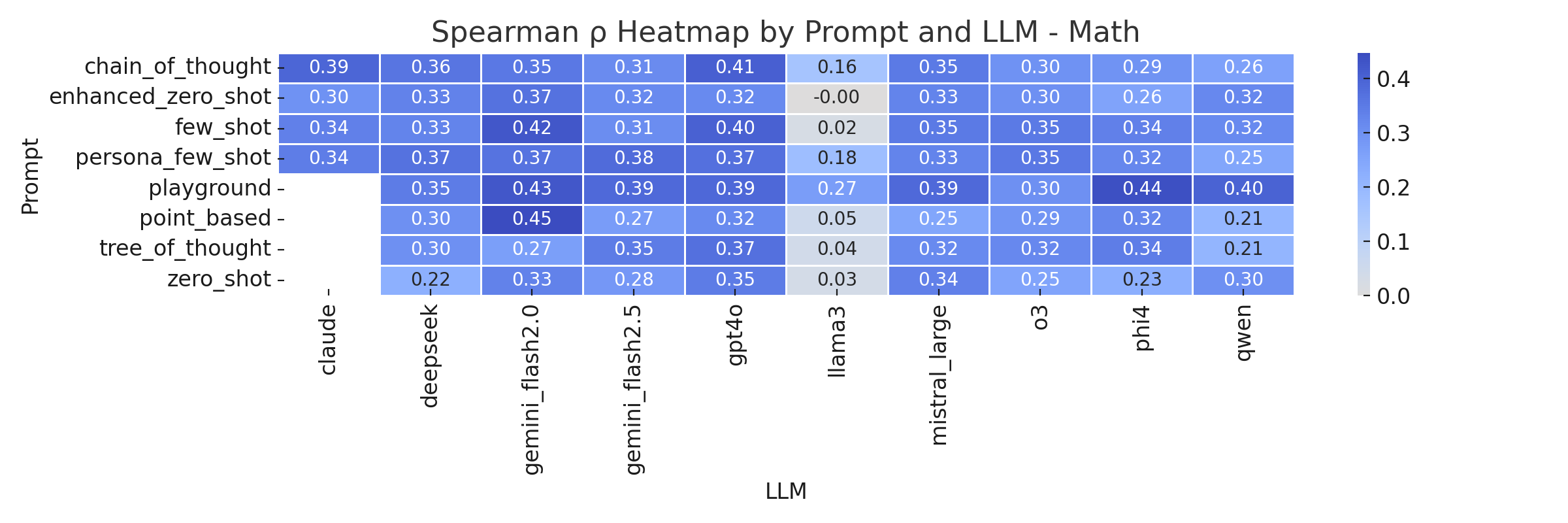}
  \caption{Mathematics: Spearman~\(\rho\) heatmap by prompt and model.}
  \label{fig:math_heatmap_spearman}
\end{figure}

\clearpage
\noindent{\bf Mathematics.}
Mathematics exhibits the strongest prompt sensitivity, with absolute error and rank fidelity varying substantially across templates for several models, indicating that rubric structure can materially change difficulty estimates in STEM-style items (mean RMSE 2.73 vs.\ 2.13 overall). Prompt choice is significant (Welch ANOVA \(F_{7,41.3}=2.98,\,p=0.009\)). The detailed results
are summarized in Fig.~\ref{fig:math_heatmap_rmse} and Fig.~\ref{fig:math_heatmap_spearman}.

\noindent
\textit{Best prompt:} \texttt{point\_based} (RMSE 2.30; \(\rho=0.29\)), suggesting rubric-style step decomposition helps multi-step reasoning.

\noindent
\textit{Best model:} \textit{Phi-4} (RMSE 2.77; \(\rho=0.29\)); \textit{Gemini-2.0-Flash} attains a higher \(\rho=0.37\) but with larger error, so \textit{Phi-4} is the more reliable overall choice for Math.

\vspace{1cm}

\vspace{1cm}

\begin{figure}[H]
  \centering
  \includegraphics[width=1.1\linewidth]{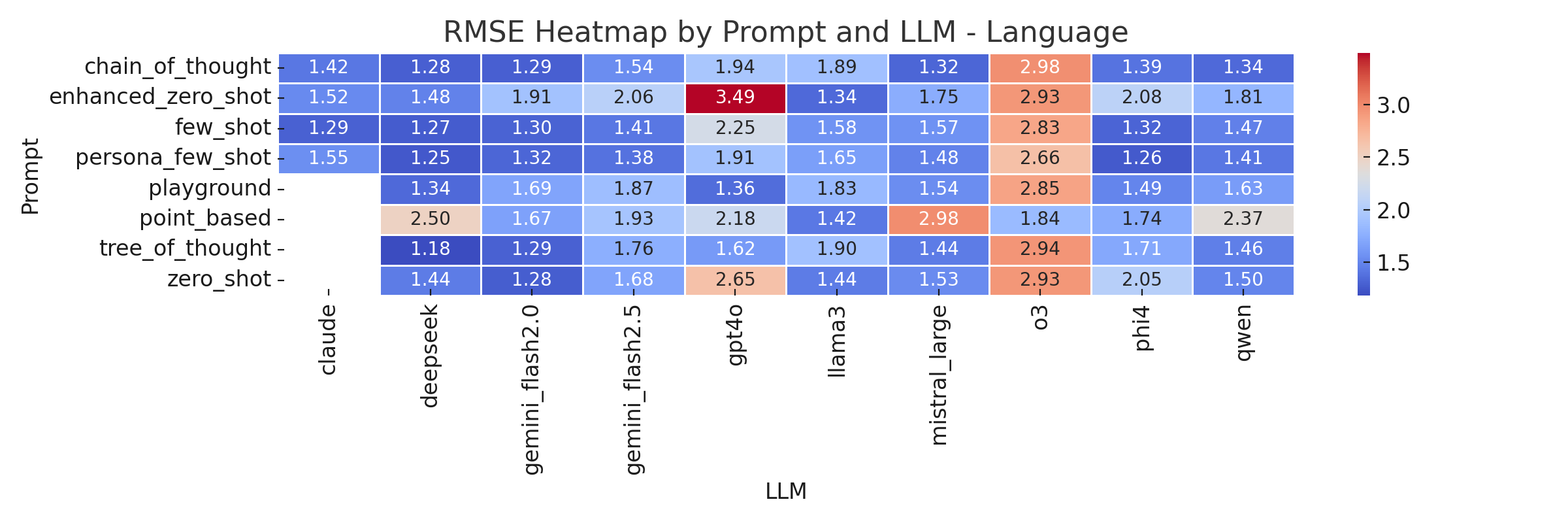}
  \caption{Languages: RMSE heatmap by prompt and model.}
  \label{fig:lang_heatmap_rmse}
\end{figure}
\begin{figure}[H]
  \centering
  \includegraphics[width=1.1\linewidth]{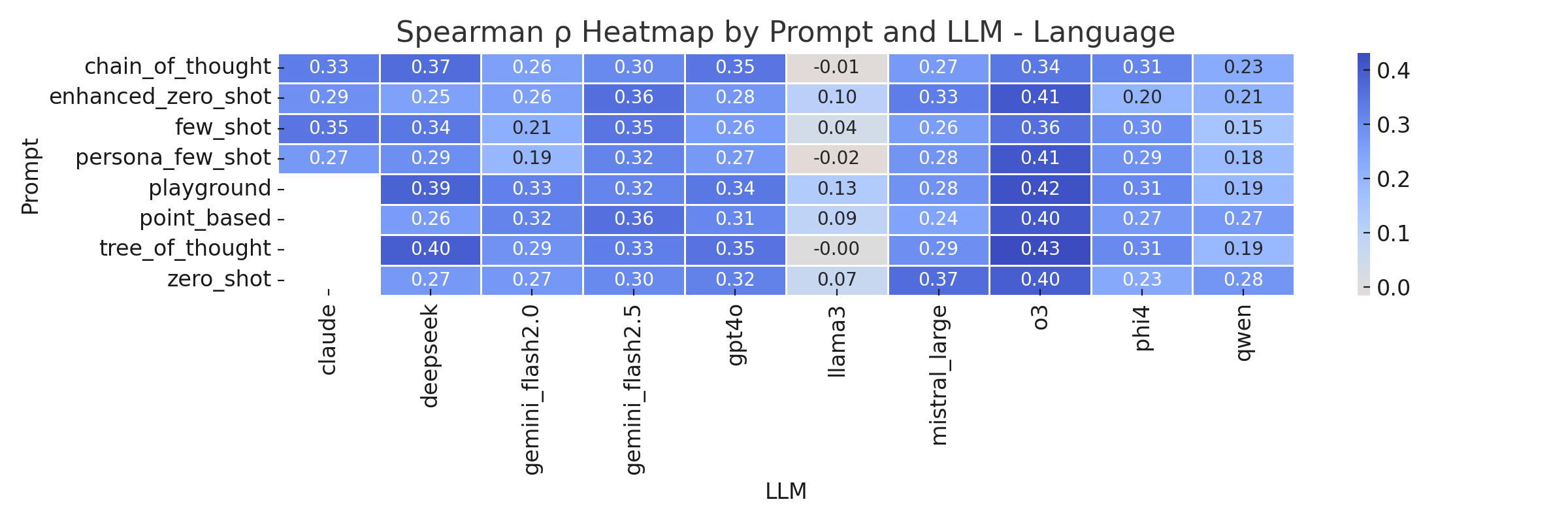}
  \caption{Languages: Spearman~\(\rho\) heatmap by prompt and model.}
  \label{fig:lang_heatmap_spearman}
\end{figure}

\clearpage
\noindent{\bf Languages.}
Languages is comparatively stable for the strongest models, but rank fidelity remains more sensitive to prompting than absolute error, suggesting that ordering is easier to disrupt than mean alignment (mean RMSE 1.72, mean \(\rho\) 0.22). The detailed results
are summarized in Fig.~\ref{fig:lang_heatmap_rmse} and Fig.~\ref{fig:lang_heatmap_spearman}.

\noindent
\textit{Best prompt:} \texttt{persona\_few\_shot} (RMSE 1.59); \texttt{playground} slightly leads in \(\rho=0.30\). Prompt effect is not significant (\(p=0.33\)).

\noindent
\textit{Best model:} \textit{Claude-4-Sonnet} (RMSE 1.44; \(\rho=0.30\)), with \textit{Phi-4} close behind—consistent with proprietary models excelling on language-heavy content.

\vspace{1cm}

\vspace{1cm}

\begin{figure}[H]
  \centering
  \includegraphics[width=1.1\linewidth]{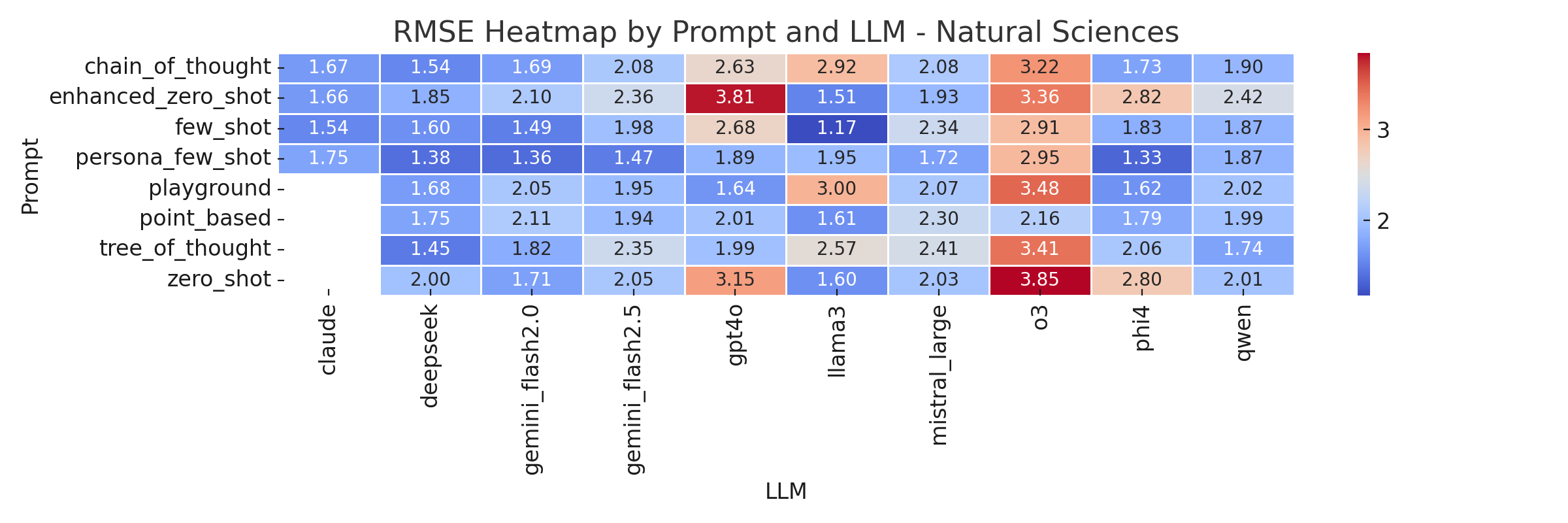}
  \caption{Natural Sciences: RMSE heatmap by prompt and model.}
  \label{fig:nat_heatmap_rmse}
\end{figure}
\begin{figure}[H]
  \centering
  \includegraphics[width=1.1\linewidth]{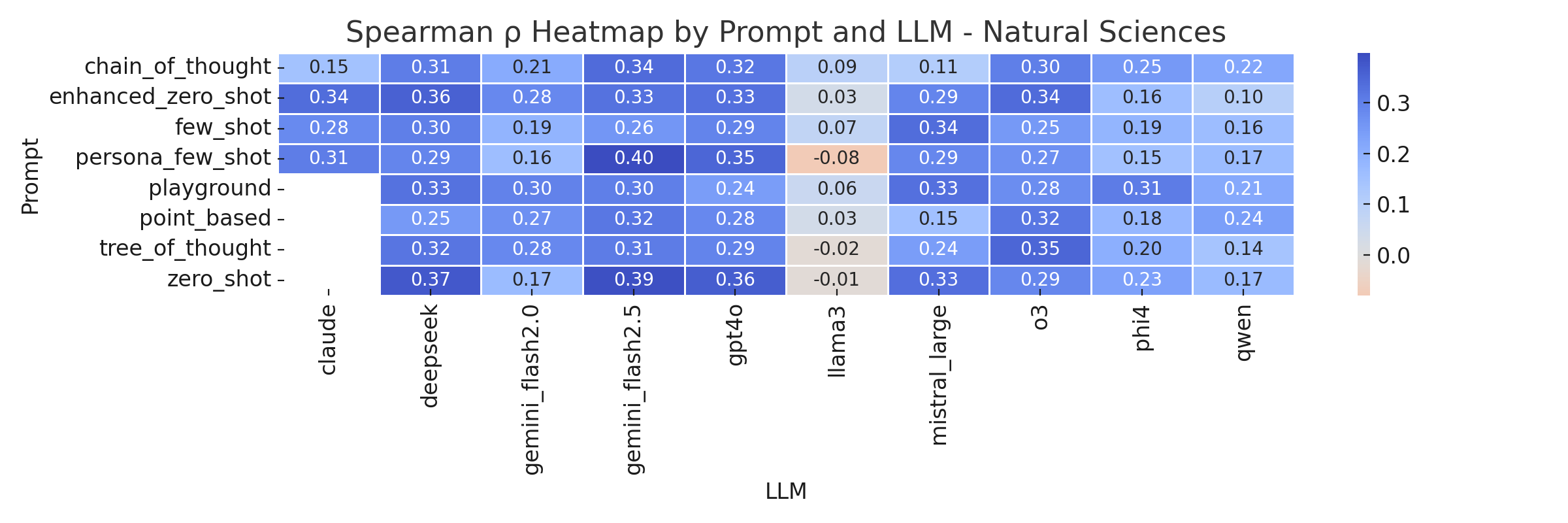}
  \caption{Natural Sciences: Spearman~\(\rho\) heatmap by prompt and model.}
  \label{fig:nat_heatmap_spearman}
\end{figure}

\clearpage
\noindent{\bf Natural Sciences.}
Natural Sciences shows a wide spread in error across prompt families and uneven rank fidelity across models, consistent with higher reasoning demands and heterogeneous item formats.
(mean RMSE 2.12); prompt differences are not significant (\(p=0.29\)).
The detailed results
are summarized in Fig.~\ref{fig:nat_heatmap_rmse} and Fig.~\ref{fig:nat_heatmap_spearman}.

\noindent
\textit{Best prompt:} \texttt{persona\_few\_shot} (RMSE 1.77); \texttt{playground} yields the best \(\rho=0.26\).

\noindent
\textit{Best model:} \textit{Claude-4-Sonnet} has the best RMSE (1.66); \textit{Gemini-2.5-Flash} attains the highest \(\rho=0.33\), again showing an error–rank trade-off.

\vspace{1cm}

\vspace{1cm}

\begin{figure}[H]
  \centering
  \includegraphics[width=1.1\linewidth]{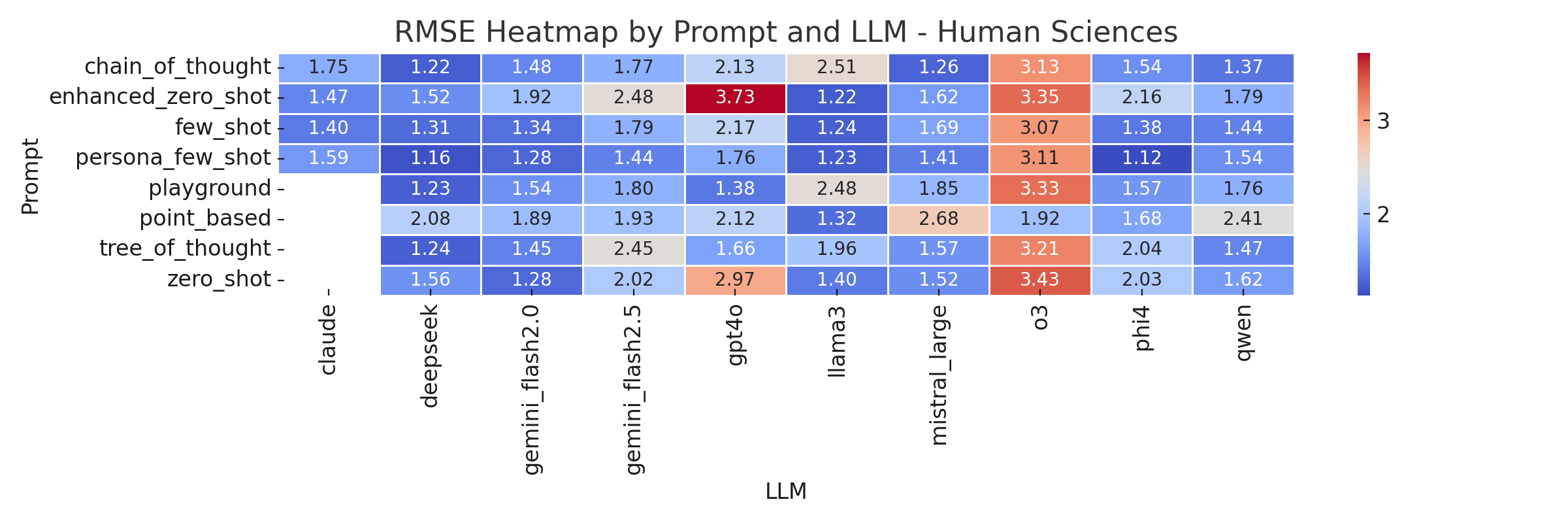}
  \caption{Human Sciences: RMSE heatmap by prompt and model.}
  \label{fig:hum_heatmap_rmse}
\end{figure}
\begin{figure}[H]
  \centering
  \includegraphics[width=1.1\linewidth]{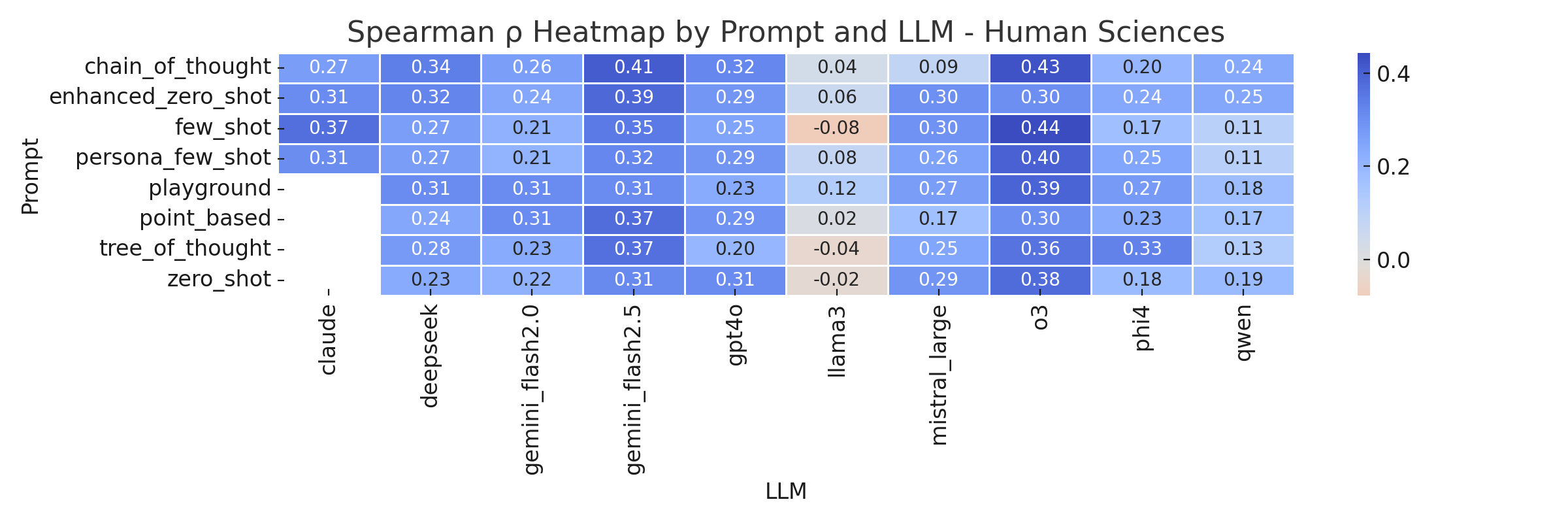}
  \caption{Human Sciences: Spearman~\(\rho\) heatmap by prompt and model.}
  \label{fig:hum_heatmap_spearman}
\end{figure}

\clearpage
\noindent{\bf Human Sciences.}
Human Sciences tends to yield lower average error for multiple model--prompt pairs, but rank fidelity still varies noticeably across prompts, so ``easy'' RMSE does not guarantee reliable ordering.
(mean RMSE 1.93).
The detailed results
are summarized in Fig.~\ref{fig:hum_heatmap_rmse} and Fig.~\ref{fig:hum_heatmap_spearman}.

\noindent
\textit{Best prompt:} \texttt{persona\_few\_shot} minimises RMSE (1.56); \texttt{enhanced\_zero\_shot} narrowly leads in \(\rho=0.27\).

\noindent
\textit{Best model:} \textit{DeepSeek-R1} (RMSE 1.41; \(\rho=0.27\)), likely reflecting benefits of reasoning-centric pretraining for socio-historical context.

\vspace{1cm}

\vspace{1cm}

\clearpage
\noindent{\bf Top Configurations by Subject and Metric.}
Table~\ref{tab:best_pair_rmse} and Table~\ref{tab:best_pair_rho} summarize the
best-performing model–prompt pairs for each subject area. The first table
focuses on configurations achieving the lowest RMSE, while the second highlights
those with the highest Spearman~\(\rho\) correlations.

\begin{table}[H]
  \centering
  \caption{Best model–prompt pair per subject for \textbf{lowest RMSE}.}
  \label{tab:best_pair_rmse}
  \begin{tabular}{l l l c}
    \toprule
    Subject & Model & Prompt & RMSE $\downarrow$ \\
    \midrule
    Human Sciences     & \textit{Phi-4}         & persona\_few\_shot & 1.1185 \\
    Natural Sciences   & \textit{Llama-3}       & few\_shot          & 1.1667 \\
    Languages          & \textit{Deepseek-R1}   & tree\_of\_thoughts & 1.1836 \\
    Mathematics        & \textit{Deepseek-R1}   & point\_based       & 1.5025 \\
    \bottomrule
  \end{tabular}
\end{table}

\begin{table}[H]
  \centering
  \caption{Best model–prompt pair per subject for \textbf{highest Spearman's~$\rho$}.}
  \label{tab:best_pair_rho}
  \begin{tabular}{l l l c}
    \toprule
    Subject & Model & Prompt & Spearman~$\rho$ $\uparrow$ \\
    \midrule
    Human Sciences     & \textit{o3}               & few\_shot          & 0.4434 \\
    Natural Sciences   & \textit{Gemini-2.5-Flash} & persona\_few\_shot & 0.3987 \\
    Languages          & \textit{o3}               & tree\_of\_thoughts & 0.4315 \\
    Mathematics        & \textit{Gemini-2.0-Flash} & point\_based       & 0.4481 \\
    \bottomrule
  \end{tabular}
\end{table}

\smallskip
\noindent{\bf Key Takeaways across Subjects.}
Based on the above resultsa and analysis,
we summarize the following key takeawys.
\begin{itemize}
  \item \textbf{\texttt{persona\_few\_shot} is the most robust prompt:}
        top-1 or top-2 RMSE in three of four subjects.
  \item \textbf{\texttt{playground} often maximises rank fidelity:}
        highest mean \(\rho\) in Mathematics, Natural Sciences, and Languages, likely due to its structured JSON output.
  \item \textbf{No single model dominates:}
        \textit{Claude-4-Sonnet} and \textit{DeepSeek-R1} lead on language-centric content; \textit{Phi-4} is the only open-weight model winning a subject (Math) on RMSE.
  \item \textbf{Mathematics is an outlier:}
        only domain with a significant prompt effect (\(p<0.01\)) and the highest overall error.
  \item \textbf{Reasoning scaffolds help multi-step items:}
        the advantage of \texttt{point\_based} and \texttt{persona\_few\_shot} in Math points to the value of explicit step decomposition.
\end{itemize}
These results show that subject matter and prompt design jointly shape accuracy and rank fidelity, and that “best” choices are objective- and domain-dependent.

\subsection{Modality Effect}
\label{sec:results_modality}

Next, we split the item set by modality (non-visual vs.\ visual, with $N_{\text{visual}}=393$ items and $N_{\text{non-visual}}=638$), keeping the model and prompt fixed and comparing performance on each subset.
Many ENEM questions contain diagrams, graphs, or tables, which we converted into textual descriptions. We investigate whether replacing visuals with text systematically impacts absolute error or rank fidelity. Throughout this section, positive residuals ($r=\hat{y}-y>0$) indicate overestimation of IRT difficulty (predicting items as harder), and negative residuals indicate underestimation.

\smallskip
\noindent{\bf Global Comparison.}
Across all 76 model–prompt pairs, visual items incur a consistent modality penalty: performance degrades relative to non-visual items under the same prompt, consistent with information loss from text-only transcription of diagrams and tables \textit{$\Delta$RMSE} (no-media $-$ media) $=\,$\textbf{$-1.26$} (SD$\approx0.95$); \textit{56/76} pairs show a significant RMSE rise ($p<0.05$).
Mean $\Delta\rho=-0.05$; ten pairs drop significantly, with the largest effects clustered in \textit{Phi-4}.
Overall, the adverse effect on error is moderate (Hedge's $g=-0.17$).

\begin{figure}[!t]
  \centering
  \includegraphics[width=\linewidth]{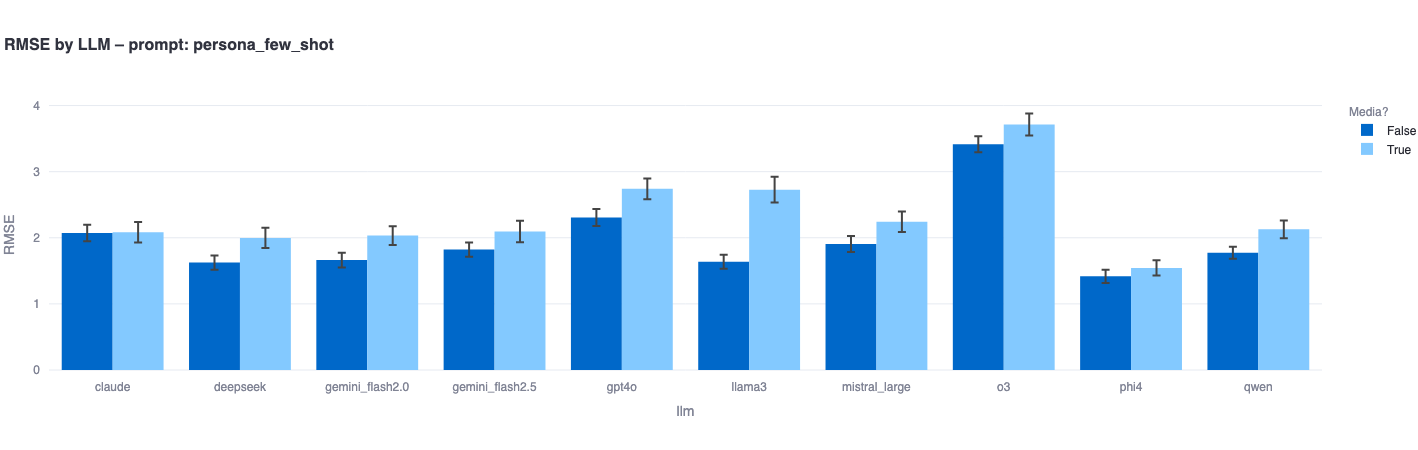}
  \caption{RMSE by model for the \texttt{persona\_few\_shot} prompt, split by modality. Error bars: 95\% CI.}
  \label{fig:media_bar_pfs}
\end{figure}

\begin{figure}[!t]
  \centering
  \includegraphics[width=\linewidth]{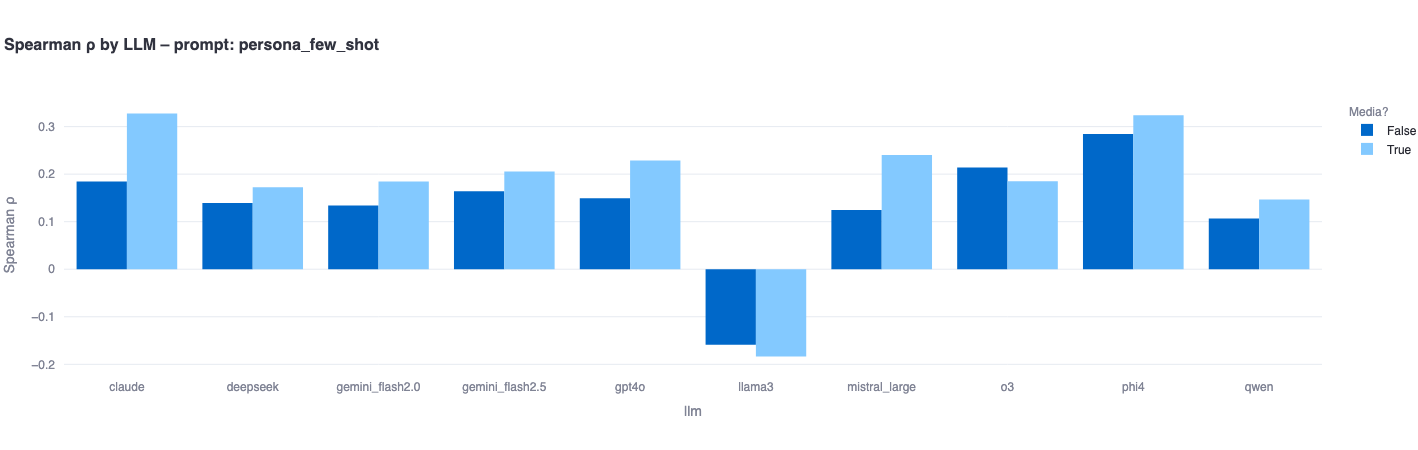}
  \caption{Spearman~$\rho$ by model for the \texttt{persona\_few\_shot} prompt, split by modality. Error bars: 95\% CI.}
  \label{fig:media_bar_spearman_pfs}
\end{figure}

\smallskip
\noindent{\bf Prompt and Model Sensitivity.}
We further examine how sensitive the model and prompts are to modality effects.
For example, 
Figure~\ref{fig:media_bar_pfs} shows the pattern under \texttt{persona\_few\_shot}: all models incur higher RMSE on visual items (from $+0.01$ for \textit{Claude-4-Sonnet} to $+1.09$ for \textit{Llama-3}). 
For the results in Fig.~\ref{fig:media_bar_spearman_pfs},
we can seee that 
\texttt{point\_based} and \texttt{enhanced\_zero\_shot} largely preserve rank fidelity and show the smallest media penalty ($|\Delta\text{RMSE}|<0.4$),
while
\texttt{persona\_few\_shot}, \texttt{playground}, and \texttt{chain\_of\_thought} often degrade with visuals.
\textit{DeepSeek-R1} is least sensitive (median $|\Delta\text{MSE}|=0.24$; no prompt significant at $p<0.05$), while \textit{o3} degrades most (median $|\Delta\text{MSE}|=0.83$; all prompts significant).

\smallskip
\noindent{\bf Alignment with Human Difficulty Trends.}
Table~\ref{tab:llm_media_gap}
Models mirror students in CN — visual items are harder, and models even overestimate this effect, but fail to capture the Math pattern.
From the results in Table~\ref{tab:llm_media_gap},
we can see that humans find visuals easier ($g_\text{human}\approx-0.39$), whereas models are essentially neutral ($g_\text{model}\approx+0.01$). In LC and CH, models show small easiness effects where humans are near zero.

\begin{table}[!h]
\centering
\caption{Mean difference (\,$\Delta$) and Hedge's $g$ for visual vs.\ non-visual questions, averaged across models and prompts. Positive $=$ models rate visuals as harder.}
\label{tab:llm_media_gap}
\begin{tabular}{lcc}
\toprule
Subject & Mean $\Delta$ & Mean $g$ \\
\midrule
Human Sciences (CH)    & $-0.22$ & $-0.18$ \\
Natural Sciences (CN)  & $+0.53$ & $+0.44$ \\
Languages (LC)         & $-0.20$ & $-0.18$ \\
Mathematics (MT)       & $+0.02$ & $+0.01$ \\
\bottomrule
\end{tabular}
\end{table}

\smallskip
\noindent{\bf Practical mitigation.}
For items involving visuals, concise, criterion-driven prompts \newline (\texttt{point\_based}, \texttt{enhanced\_zero\_shot}) are preferable, as lengthy reasoning templates introduce brittleness. Direct multimodal ingestion of raw images represents a promising avenue for future work but is not assessed in this study.

\subsection{Prompt Sensitivity and Stability}
\label{sec:prompt-stability}

We quantify how much predictions vary across eight prompt templates (see details in Section~\ref{sec:prompts}) using two statistics per model $m$:
\begin{align}
\sigma_{\text{prompt}}^{(m)} &=
  \operatorname*{median}_{q_i}
  \Bigl[\operatorname{SD}\bigl(y_{i1},\dots,y_{i8}\bigr)\Bigr], \\
\mathrm{KL}_{\text{mean}}^{(m)} &=
  \tfrac18 \sum_{j=1}^{8}
  \mathrm{KL}\!\bigl(f_{m,p_j}\,\|\,f_{\textsc{irt}}\bigr),
\end{align}
where $y_{ij}$ is the predicted difficulty of question $q_i$ under prompt $p_j$, and $f_{m,p_j}$ / $f_{\textsc{irt}}$ are Kernel Density Estimates of the model/IRT distributions.
KL is the Kullback-Leibler (KL) Divergence function that quantifies how one probability distribution differs from a second.
Lower $\sigma_{\text{prompt}}$ means higher prompt stability; lower $\mathrm{KL}_{\text{mean}}$ indicates better distributional calibration. From the results presented in Table~\ref{tab:prompt-stability-alt}, we can find that \textit{Claude-4-Sonnet} and \textit{Gemini-2.0-Flash} show tightly stacked prompt curves ($\sigma_{\text{prompt}}<0.65$). \textit{GPT-4o} is most prompt-sensitive ($\sigma_{\text{prompt}}=1.60$).
\textit{DeepSeek-R1} best matches the IRT distribution (lowest $\mathrm{KL}_{\text{mean}}$). \textit{o3} reproduces the IRT shape but is shifted $\approx$2–3 points too easy (worst KL).
Stable families (\texttt{zero\_shot}, \texttt{enhanced\_zero\_shot}, \texttt{few\_shot}, \texttt{persona\_few\_shot}) reduce spread (–10–25\%) but also compress tails; and \texttt{chain\_of\_thought}/\texttt{tree\_of\_thoughts} increase spread (+20–40\%) and sometimes improve right-tail coverage; \texttt{point\_based} uniquely populates the hardest bins ($\ge$9) at the cost of extra variance.

\begin{table}[H]
  \small
  \centering
  \caption{Prompt stability ($\sigma_{\text{prompt}}$) and calibration ($\mathrm{KL}_{\text{mean}}$). Bold = best, underline = worst.}
  \label{tab:prompt-stability-alt}
  \begin{tabularx}{\linewidth}{@{}l r c r c X@{}}
    \toprule
    \textbf{Model} & $\boldsymbol{\sigma_{\text{prompt}}}$ & \textbf{Rank$_{\sigma}$} & $\boldsymbol{\mathrm{KL}_{\text{mean}}}$ & \textbf{Rank$_{\mathrm{KL}}$} & \textbf{Verdict} \\
    \midrule
    Claude-4-Sonnet & \textbf{0.577} & \textbf{1} & 0.701 & 4 & Most stable; mild hard-tail bias. \\
    Gemini-2.0-Flash & 0.641 & 2 & 0.712 & 5 & Stable; close to IRT shape. \\
    DeepSeek-R1 & 1.035 & 3 & \textbf{0.441} & \textbf{1} & Best distributional match. \\
    Qwen-3 & 1.035 & 4 & 0.626 & 3 & Moderate spread; compressed tails. \\
    o3 & 1.035 & 5 & \underline{3.345} & \underline{10} & Stable shape, large left shift. \\
    Gemini-2.5-Flash & 1.165 & 6 & 0.789 & 6 & More variance; minor calibration gain. \\
    Phi-4 & 1.126 & 7 & 0.722 & 7 & Twin modes at 4 and 7. \\
    Mistral-Large & 1.302 & 8 & 0.762 & 8 & Rubric inflates hard tail. \\
    Llama-3 & 1.356 & 9 & 0.827 & 9 & Hard-biased modes (7–8). \\
    GPT-4o & \underline{1.598} & \underline{10} & 1.603 & 2 & Largest prompt spread. \\
    \bottomrule
  \end{tabularx}
\end{table}

\subsection{Error Landscape}
\label{sec:qual_landscape}

\begin{figure}[!t]
  \centering
  \includegraphics[width=1\linewidth]{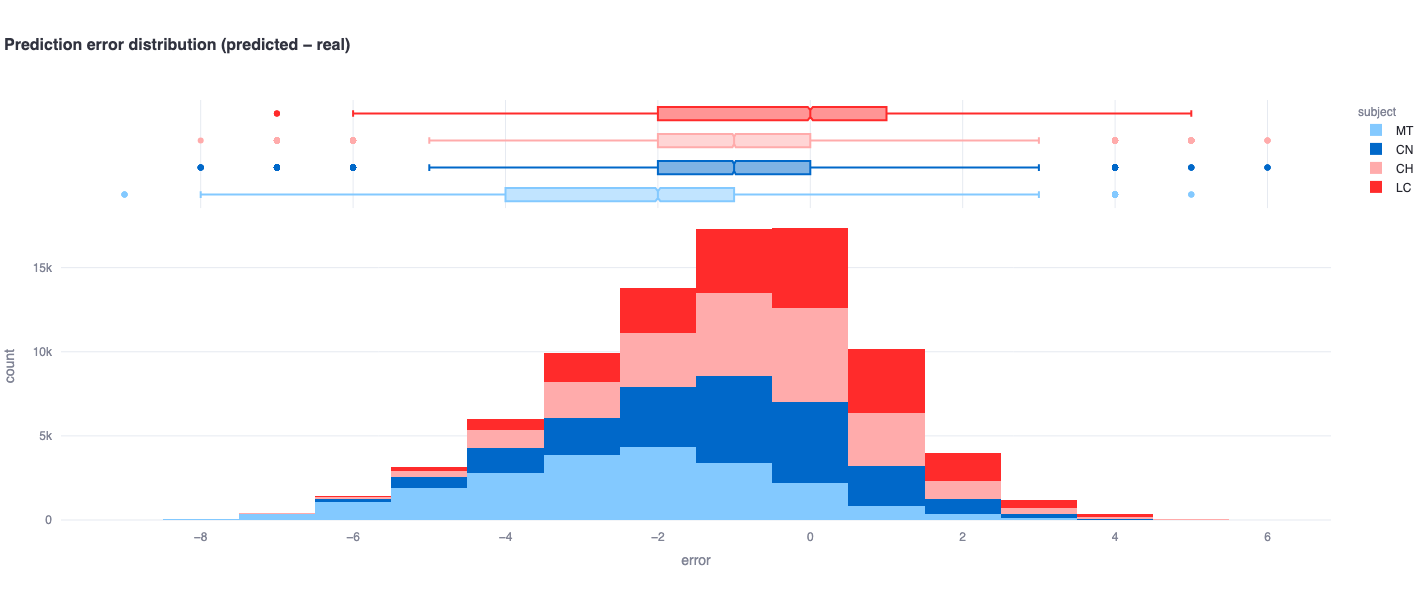}
  \caption{Signed residuals $\Delta=\mathrm{prediction}-\mathrm{IRT}$ for all
           extreme cases, colour-coded by subject.}
  \label{fig:error_by_subject}
\end{figure}

Fig.~\ref{fig:error_by_subject} shows three stable patterns:
(i) \emph{STEM underestimation} -- Mathematics and Natural Sciences account for most large negative residuals; (ii) \emph{LC inflation}, Languages \& Codes contributes disproportionately to the positive tail; (iii) \emph{Human Sciences balance} -- residuals concentrate near~0. These signatures align with the heavier use of diagrams in STEM and the rubric emphasis on stylistic nuance in LC.

\begin{table}[!t]
  \centering
  \caption{Extreme residuals ($|\Delta|>3$) by model. Negative residuals dominate.}
  \label{tab:extreme-models}
  \small
  \begin{tabular}{lrrrrrr}
    \toprule
    Model & \# extremes & Share (\%) & \#\,$\Delta$<0 & \#\,$\Delta$>0 & \%\,$\Delta$<0 & Mean $\Delta$ \\
    \midrule
    o3                      & 3,232 & 28.4 & 3,230 & 2   & 99.9 & -4.79 \\
    GPT-4o                  & 1,896 & 16.7 & 1,851 & 45  & 97.6 & -4.46 \\
    Gemini-2.5-Flash        & 1,029 &  9.0 & 1,003 & 26  & 97.5 & -4.41 \\
    Llama-3                 & 1,000 &  8.8 &   979 & 21  & 97.9 & -4.49 \\
    Mistral-Large           &   941 &  8.3 &   792 & 149 & 84.2 & -3.24 \\
    Phi-4                   &   895 &  7.9 &   875 & 20  & 97.8 & -4.42 \\
    Gemini-2.0-Flash        &   662 &  5.8 &   660 & 2   & 99.7 & -4.47 \\
    DeepSeek-R1             &   639 &  5.6 &   576 & 63  & 90.1 & -3.77 \\
    Qwen-3                  &   625 &  5.5 &   561 & 64  & 89.8 & -3.51 \\
    Claude-4-Sonnet         &   449 &  3.9 &   449 & 0   & 100.0 & -4.69 \\
    \midrule
    \textbf{Total}          & 11,368 & 100.0 & 10,476 & 892 & 92.2 & -4.29 \\
    \bottomrule
  \end{tabular}
\end{table}

Table~\ref{tab:extreme-models} shows the extreme residuals for each model.
Two mechanisms explain most extremes: (a) uniform calibration offsets (near-purely negative tails; e.g., \textit{o3}, Gemini, Phi-4, Llama-3), suggesting a global left shift correctable via post-hoc calibration (see Section~\ref{sec:results-calibration}); and (b) isolated reasoning blow-ups (non-trivial positive tails; e.g., GPT-4o, Mistral, DeepSeek-R1, Qwen-3) on multi-step STEM or nuance-heavy LC items, which likely require prompt/inputs fixes rather than rescaling.

\subsection{Post Hoc Calibration of Model Outputs}
\label{sec:results-calibration}

All models underestimated item difficulty by a nearly uniform offset. The mean residual \(r = y - \hat{y}\) was positive for every system, largest for \textit{o3} (\(\approx+3\) points), followed by \textit{GPT-4o} (\(\approx+1.8\)), and clustered near \(+1\) for \textit{Gemini-2.0-Flash}, \textit{Claude-4-Sonnet}, \textit{Phi-4}, \textit{Llama-3}, \textit{DeepSeek-R1}, \textit{Qwen-3}, and \textit{Mistral-Large}. Thus, a substantial share of error is a location bias rather than mis-ranking.

We applied the shift-only procedure with cross-validated folds and recomputed metrics. Table~\ref{tab:calib-shift} reports calibrated scores and deltas (\(\Delta\)=calibrated–raw). As expected, Spearman's \(\rho\) is unchanged for all models (ordering preserved), while RMSE/MAE improve wherever a strong offset existed.

\begin{table}[H]
  \centering
  \caption{Shift-only calibrated metrics per model (first row) and change vs.\ raw (second row). $\Delta$= calibrated $-$ raw.}
  \label{tab:calib-shift}
  \setlength{\tabcolsep}{5pt}
  \begin{tabular}{lcccccc}
    \toprule
    Model & RMSE$\downarrow$ & MAE$\downarrow$ & W1A$\uparrow$ & EM$\uparrow$ & $\rho$$\uparrow$ \\
    \midrule
    o3              & 1.801 & 1.405 & 0.431 & 0.235 & 0.209 \\
    $\Delta$        & $-1.733$ & $-1.696$ & $+0.236$ & $+0.175$ & $0.000$ \\
    \addlinespace[2pt]
    GPT-4o          & 2.172 & 1.784 & 0.316 & 0.160 & 0.158 \\
    $\Delta$        & $-0.621$ & $-0.465$ & $-0.076$ & $+0.012$ & $0.000$ \\
    \addlinespace[2pt]
    Gemini-2.5-Flash& 1.956 & 1.565 & 0.371 & 0.195 & 0.244 \\
    $\Delta$        & $-0.326$ & $-0.188$ & $-0.145$ & $+0.003$ & $0.000$ \\
    \addlinespace[2pt]
    Gemini-2.0-Flash& 1.574 & 1.258 & 0.464 & 0.261 & 0.167 \\
    $\Delta$        & $-0.454$ & $-0.301$ & $-0.101$ & $+0.051$ & $0.000$ \\
    \addlinespace[2pt]
    Claude-4-Sonnet & 1.748 & 1.379 & 0.409 & 0.240 & 0.191 \\
    $\Delta$        & $-0.380$ & $-0.183$ & $-0.182$ & $-0.003$ & $0.000$ \\
    \addlinespace[2pt]
    Phi-4           & 1.870 & 1.462 & 0.361 & 0.208 & 0.172 \\
    $\Delta$        & $-0.265$ & $-0.130$ & $-0.214$ & $-0.020$ & $0.000$ \\
    \addlinespace[2pt]
    Llama-3         & 2.016 & 1.596 & 0.401 & 0.187 & -0.065 \\
    $\Delta$        & $-0.211$ & $-0.078$ & $-0.152$ & $-0.026$ & $0.000$ \\
    \addlinespace[2pt]
    Mistral-Large   & 2.063 & 1.595 & 0.401 & 0.211 & 0.119 \\
    $\Delta$        & $-0.197$ & $-0.157$ & $-0.113$ & $+0.022$ & $0.000$ \\
    \addlinespace[2pt]
    DeepSeek-R1     & 1.823 & 1.428 & 0.456 & 0.221 & 0.151 \\
    $\Delta$        & $-0.147$ & $-0.043$ & $-0.147$ & $-0.015$ & $0.000$ \\
    \addlinespace[2pt]
    Qwen-3          & 1.743 & 1.337 & 0.453 & 0.246 & 0.096 \\
    $\Delta$        & $-0.311$ & $-0.283$ & $-0.078$ & $+0.060$ & $0.000$ \\
    \bottomrule
  \end{tabular}
\end{table}

\smallskip
The shift removes large location errors (e.g., \textit{o3}: RMSE $-1.73$, MAE $-1.70$) and yields meaningful gains for \textit{GPT-4o}, \textit{Gemini-2.0-Flash}, and others, while preserving rank order (\(\rho\) unchanged). Residual differences persist by slice (STEM, visual items, hardest quartile) and in the rate of extreme errors; calibration complements rather than replaces the raw comparison.

A simple additive layer can correct biases that prompt design alone does not address. Deployments requiring absolute scores should learn and apply this shift using recent labeled data. Coefficients are domain- and scale-specific and must be revalidated whenever cohorts or scales change. Adding a slope term further reduces RMSE
by 0.3–0.9 for most models.

\section{Bias Analysis under Country Context Cues}
\label{sec:bias-analysis}

Difficulty predictions may be used in workflows that affect item selection and review. Even small context cues can shift LLM outputs in systematic ways. We therefore measure how a minimal country reference influences predicted difficulty and we report the size, direction, and stability of these shifts.

We reuse the main pipeline and prompts, changing only one token in the instruction: we add the clause \emph{“students in \{\textit{Country}\}”}. We evaluate four representative models (\textit{GPT-4o, o3, Phi-4, Llama-3}) on the same 1,031 ENEM items for seven countries (Brazil, China, Dominican Republic, Finland, Hungary, Luxembourg, Philippines). For each item $i$, model $m$, and country $c$ we define the signed residual
\[
\Delta_{i,m,c} \;=\; y_i \;-\; \hat{y}_{i,m,c},
\]
where $y_i$ is the IRT difficulty mapped to the 1–10 scale and $\hat{y}_{i,m,c}$ is the model prediction under the country cue. Positive $\Delta$ means the model underestimates true difficulty. Decoding and parsing settings are unchanged.

We quantify cue sensitivity at four complementary levels: country means, responsiveness, interaction effects, and external alignment. Together they describe both the direction and structure of any bias.

\smallskip
\noindent\textit{(1) Country means by subject.} \;
For each subject $s$, model $m$, and country $c$, we compute the mean residual
\[
\bar{\Delta}_{m,c,s} = 
\frac{1}{|I_s|} \sum_{i \in I_s} \Delta_{i,m,c},
\]
where $\Delta_{i,m,c} = y_i - \hat{y}_{i,m,c}$.
Positive values indicate that the model underestimates item difficulty relative to the IRT ground truth; negative values indicate overestimation. These means summarize systematic offsets for each model–country–subject combination.

\smallskip
\noindent\textit{(2) Responsiveness (plasticity).} \;
To capture how strongly predictions react to the country cue, we compute two dispersion metrics:
\[
\mathrm{PI}_{m,s} = 
\mathrm{SD}_c\!\left(\bar{\Delta}_{m,c,s}\right),
\qquad
\mathrm{QLV}_{m,s} = 
\frac{1}{|I_s|}
\sum_{i \in I_s} 
\mathrm{Var}_c\!\left(\Delta_{i,m,c}\right).
\]
The \textit{Plasticity Index} ($\mathrm{PI}$) measures between-country separation of average residuals, while the \textit{Quasi-Linear Variance} ($\mathrm{QLV}$) captures the average within-item variability across countries. High values on either metric indicate stronger model responsiveness to contextual tokens.

\smallskip
\noindent\textit{(3) Two-factor ANOVA.} \;
We conduct a two-way fixed-effects ANOVA on all $\Delta$ values with factors
\texttt{model} and \texttt{country}, both per subject and globally.
A significant \texttt{country} main effect implies systematic mean shifts between country tokens,
and a significant \texttt{model}$\times$\texttt{country} interaction indicates that
the direction or magnitude of the cue effect differs across models.

\smallskip
\noindent\textit{(4) External alignment (descriptive).} \;
Finally, we correlate each model's country-level means $\bar{\Delta}_{m,c,\cdot}$
with the corresponding PISA Reading ranks.
This optional analysis examines whether any residual pattern loosely aligns with population-level performance trends.
It is purely descriptive and does not justify differential treatment across countries.

\clearpage
\noindent{\bf Global Pattern.}
Adding a country cue amplifies the existing underestimation bias ($\Delta>0$), strongest in Mathematics and for \textit{GPT-4o} and \textit{o3}. Averaged over countries, model ordering by mean residual is
\textit{o3} $\ge$ \textit{GPT-4o} $\gg$ \textit{Phi-4} $\ge$ \textit{Llama-3}, and subject magnitudes follow \text{MT $\gg$ CN $\gtrsim$ CH $\ge$ LC}. Illustrative means (by subject, averaged over countries) are:
\begin{itemize}[leftmargin=1.2em]
  \item \text{MT:} \textit{GPT-4o} 3.57;\; \textit{o3} 4.40;\; \textit{Phi-4} 2.48;\; \textit{Llama-3} 1.41
  \item \text{CN:} \textit{GPT-4o} 2.97;\; \textit{o3} 3.29;\; \textit{Phi-4} 1.76;\; \textit{Llama-3} 0.94
  \item \text{CH:} \textit{GPT-4o} 2.58;\; \textit{o3} 2.48;\; \textit{Phi-4} 0.54;\; \textit{Llama-3} 0.70
  \item \text{LC:} \textit{GPT-4o} 2.08;\; \textit{o3} 1.76;\; \textit{Phi-4} 0.37;\; \textit{Llama-3} 0.28
\end{itemize}


\begin{figure}[H]
  \centering
  
  \begin{subfigure}{1.0\linewidth}
    \centering
    \includegraphics[width=\linewidth]{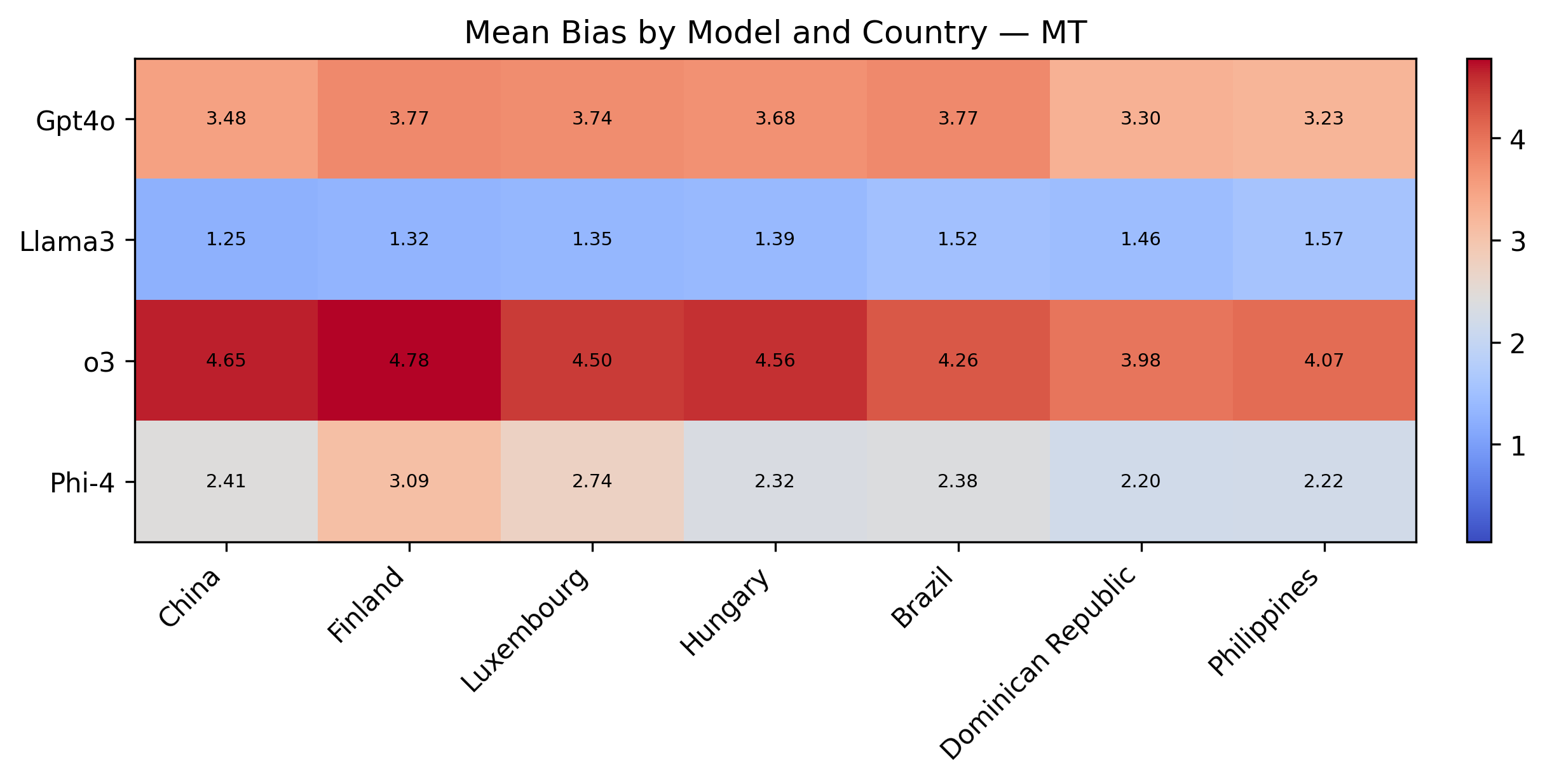}
    \caption{Mathematics}
  \end{subfigure}\\[2ex]

  \begin{subfigure}{1.0\linewidth}
    \centering
    \includegraphics[width=\linewidth]{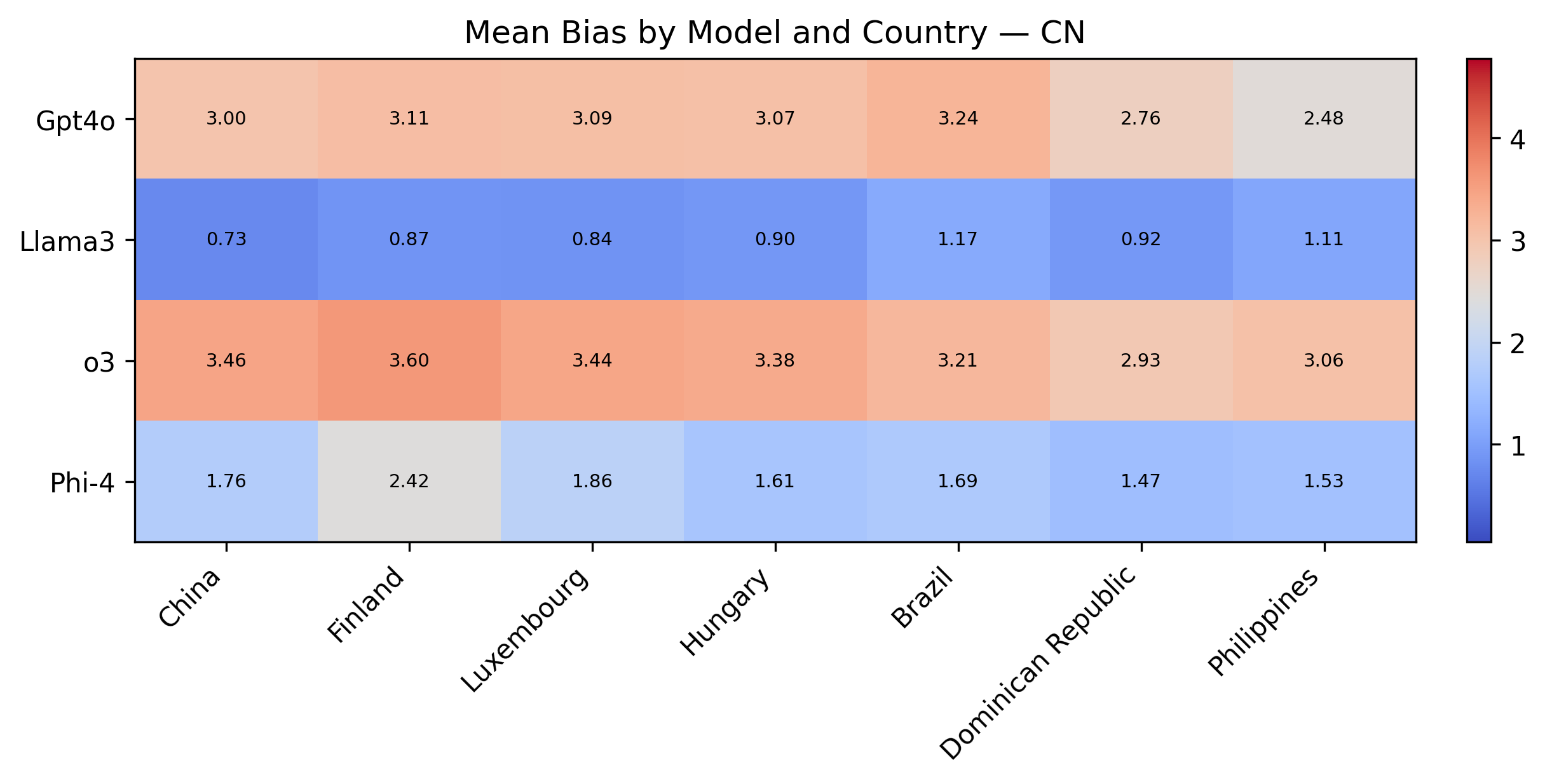}
    \caption{Natural Sciences}
  \end{subfigure}

  \caption{Mean residual $\Delta$ (IRT$-\hat{y}$) by country and model (Part 1).}
  \label{fig:bias_part1}
\end{figure}

\begin{figure}[H]
  \centering

  \begin{subfigure}{1.0\linewidth}
    \centering
    \includegraphics[width=\linewidth]{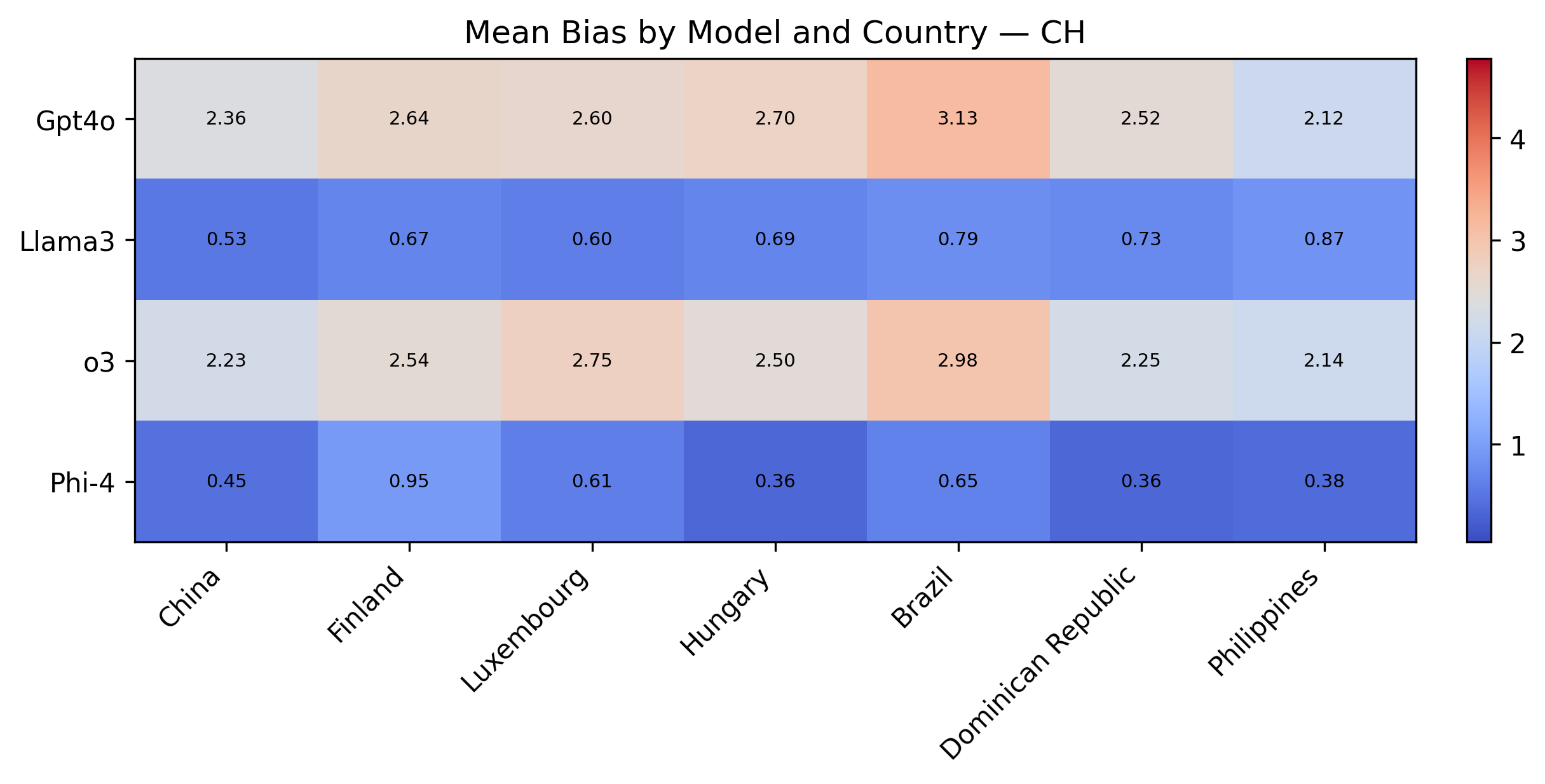}
    \caption{Human Sciences}
  \end{subfigure}\\[2ex]

  \begin{subfigure}{1.0\linewidth}
    \centering
    \includegraphics[width=\linewidth]{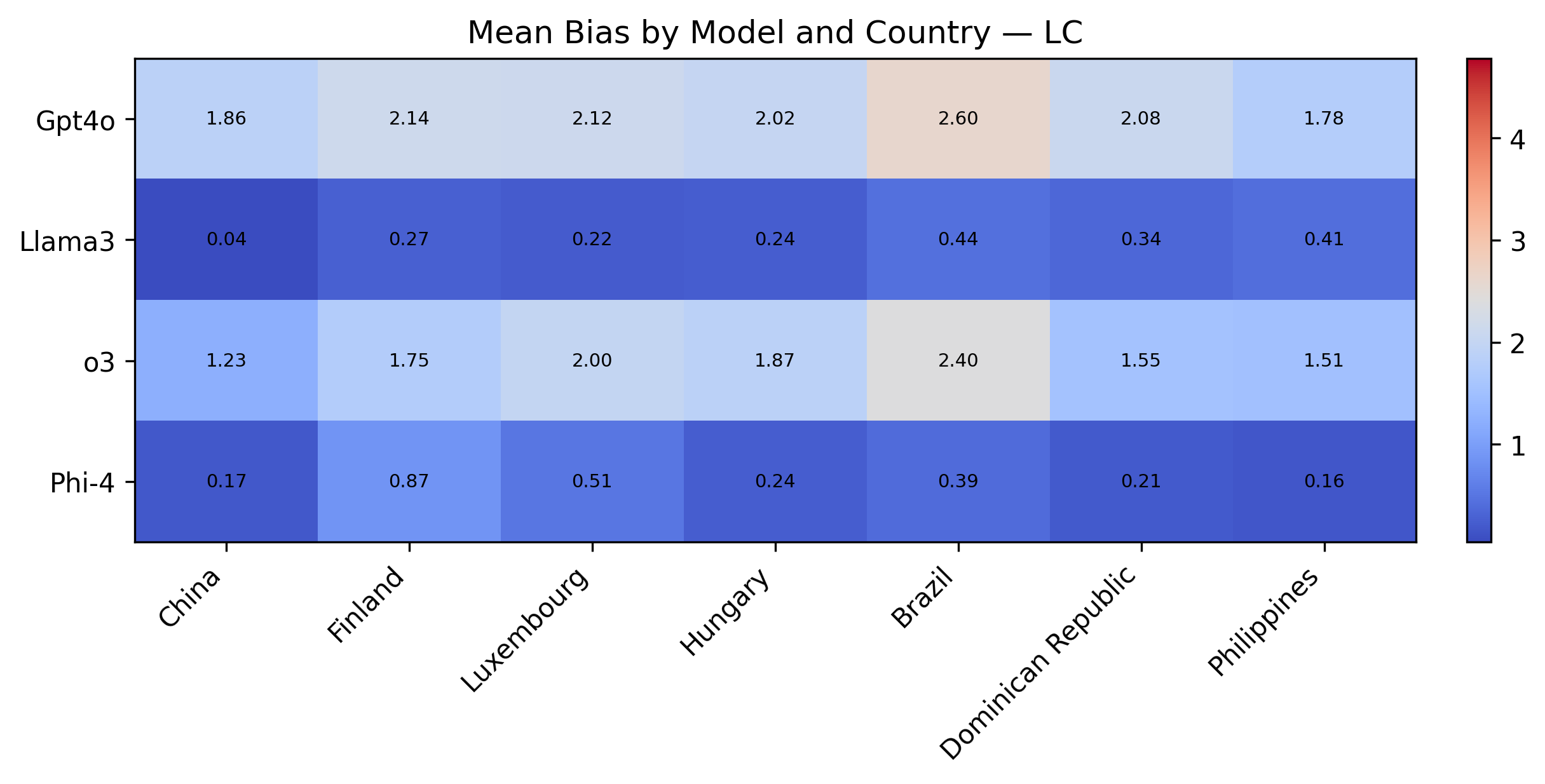}
    \caption{Languages}
  \end{subfigure}

  \caption{Mean residual $\Delta$ (IRT$-\hat{y}$) by country and model (Part 2).}
  \label{fig:bias_part2}
\end{figure}

\begin{figure}[H]
  \centering
  \begin{subfigure}{.48\linewidth}
    \centering\includegraphics[width=\linewidth]{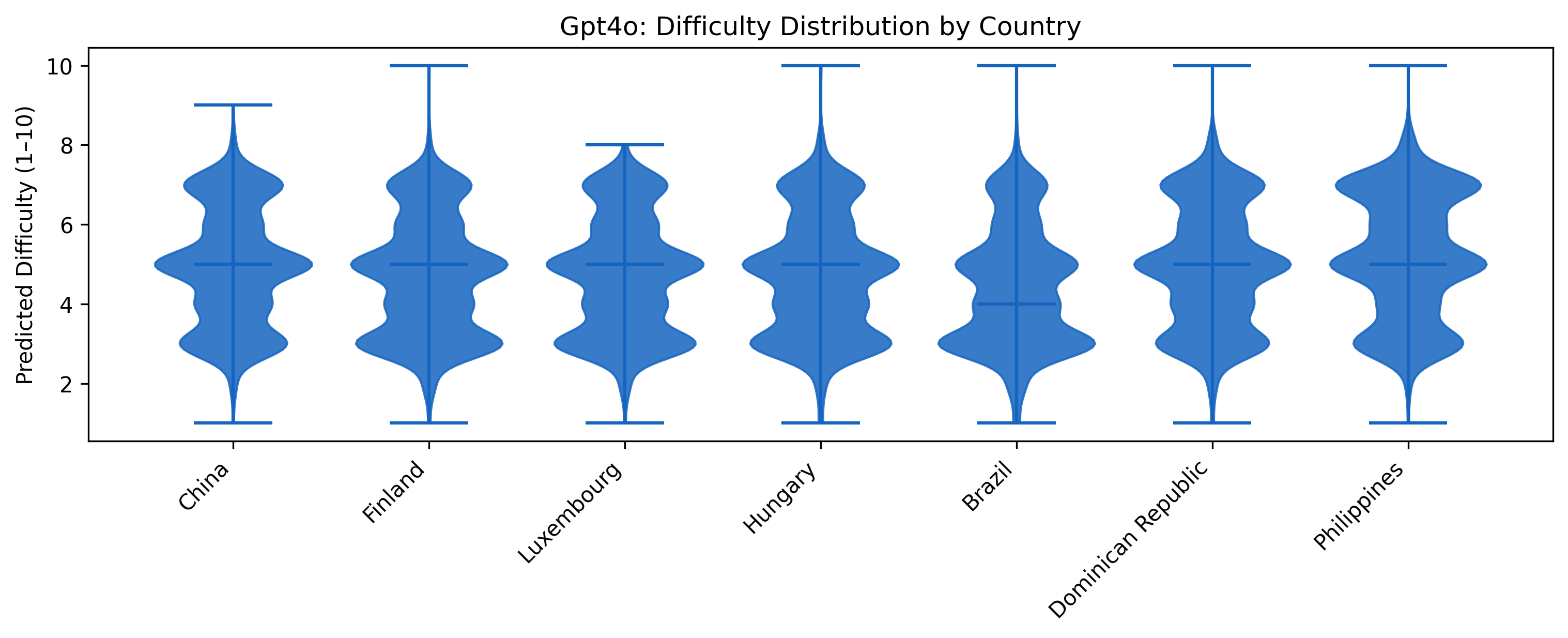}
    \caption{GPT-4o}
  \end{subfigure}\hfill
  \begin{subfigure}{.48\linewidth}
    \centering\includegraphics[width=\linewidth]{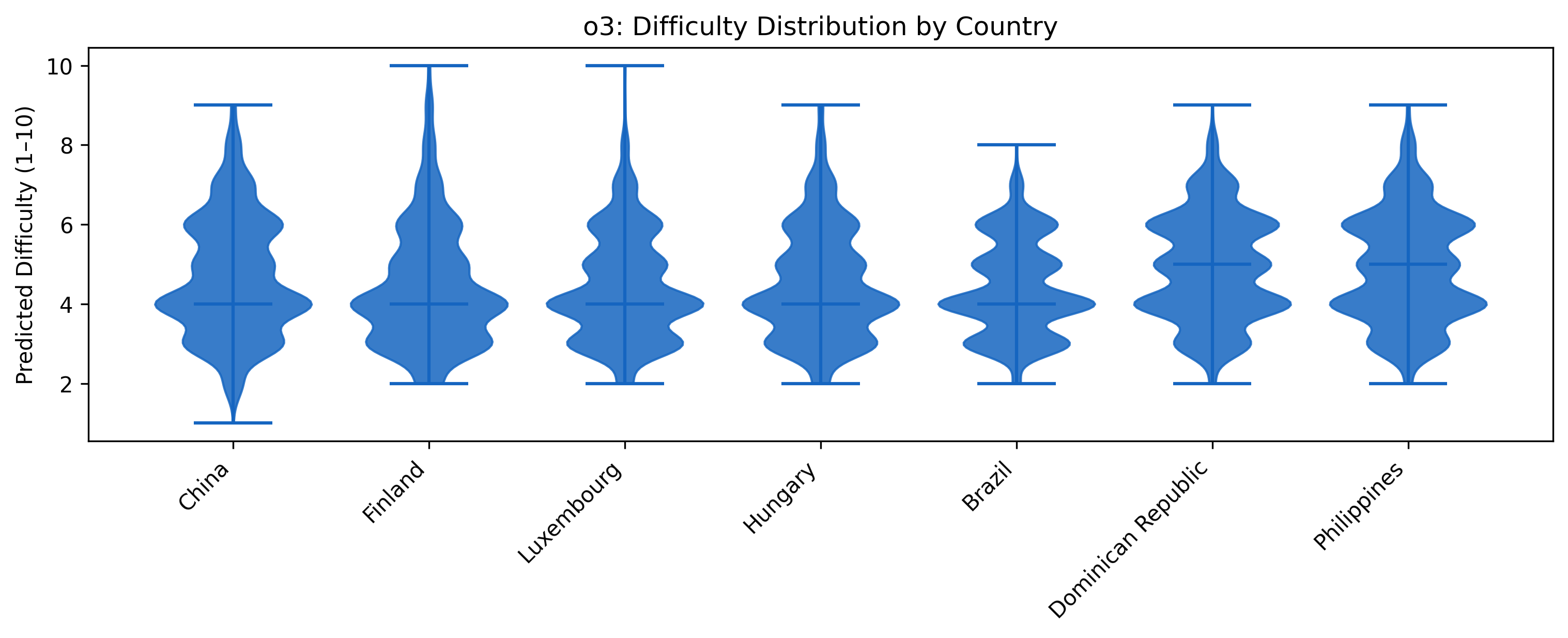}
    \caption{o3}
  \end{subfigure}\\[0.8ex]
  \begin{subfigure}{.48\linewidth}
    \centering\includegraphics[width=\linewidth]{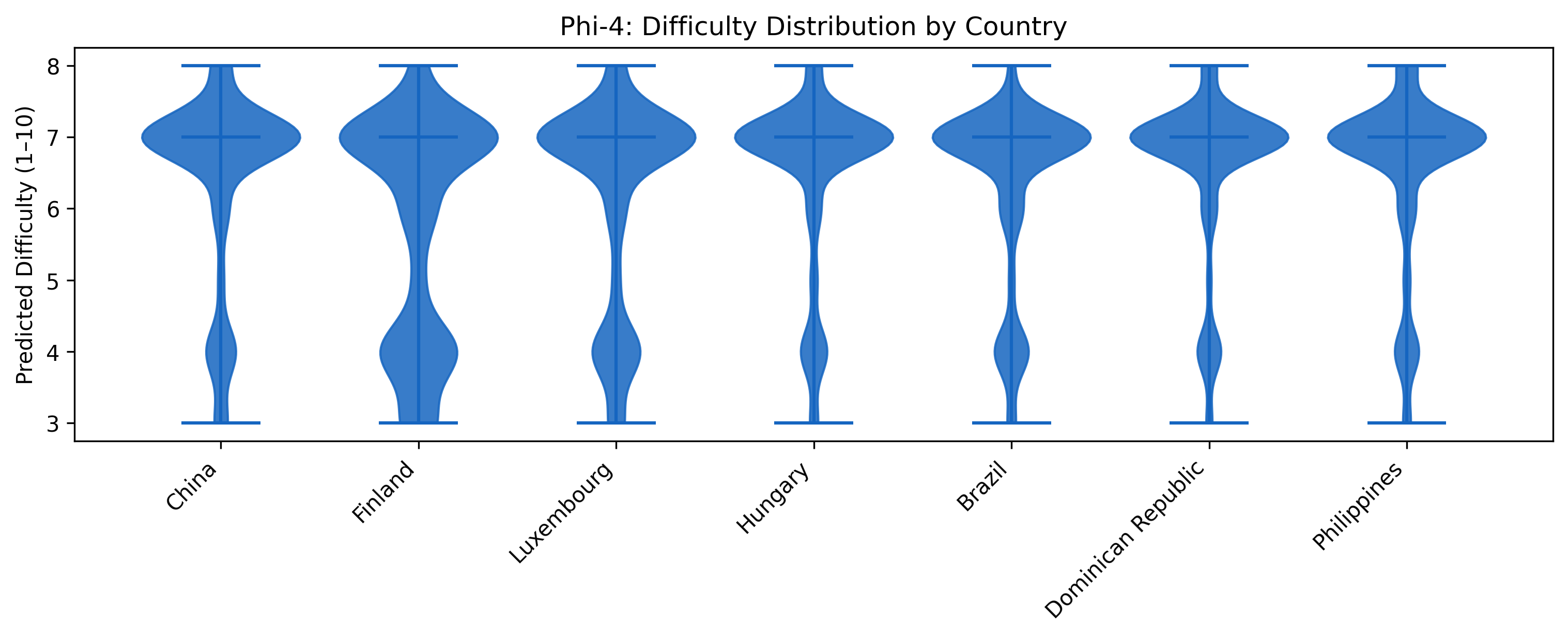}
    \caption{Phi-4}
  \end{subfigure}\hfill
  \begin{subfigure}{.48\linewidth}
    \centering\includegraphics[width=\linewidth]{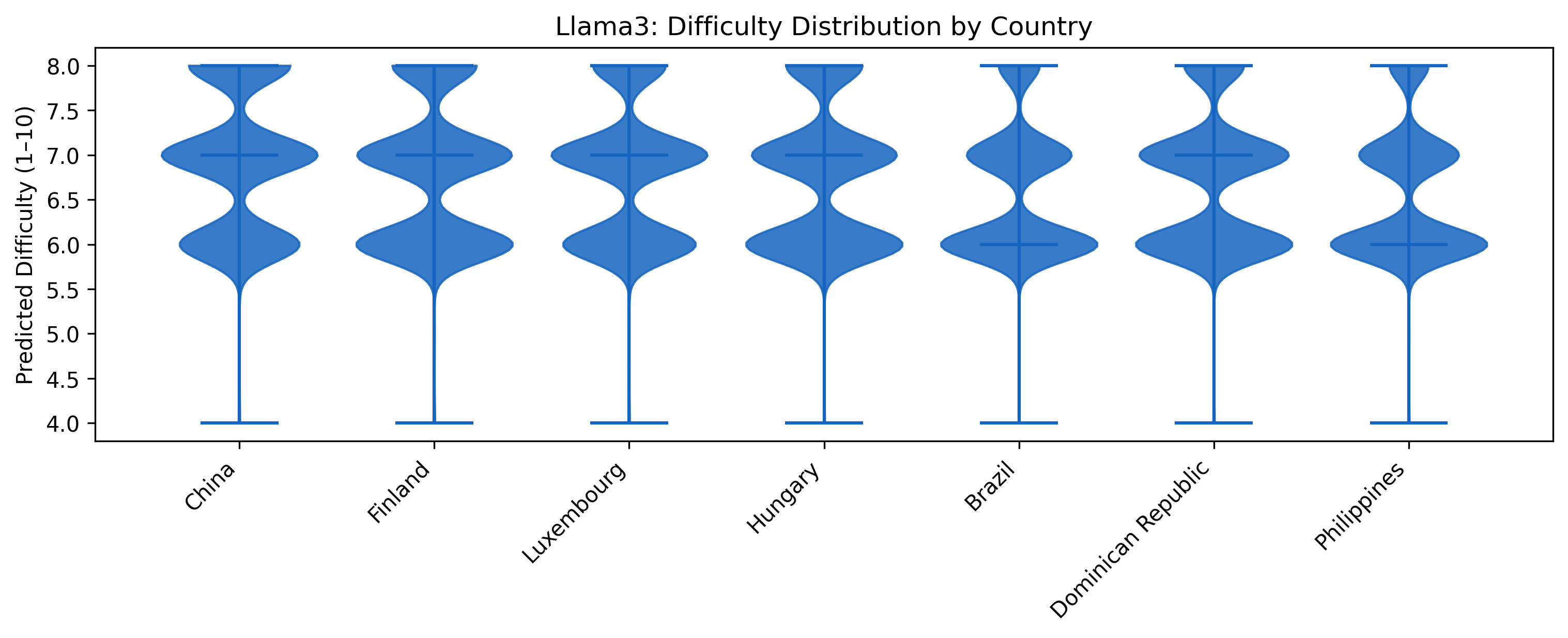}
    \caption{Llama-3}
  \end{subfigure}
  \caption{Per country distributions of predicted difficulty for each model. Similar shapes within a model indicate item wise shifts rather than large global offsets.}
  \label{fig:bias_violins}
\end{figure}

\smallskip
\noindent{\bf Country Pattern and Interactions.}
Countries differ in a model dependent way. A two factor ANOVA on $\Delta$ shows strong \textit{country} main effects and significant \textit{model}$\times$\textit{country} interactions in every subject (all $p \ll 0.001$; Table~\ref{tab:bias-anova}). Despite these differences, within model country marginals are shape similar (Fig.~\ref{fig:bias_violins}), suggesting the bias manifests primarily as item level shifts rather than large country specific global offsets.

\begin{table}[H]
  \centering
  \caption{Two factor ANOVA on $\Delta$ by subject. We report the \textit{country} main effect and the \textit{model}$\times$\textit{country} interaction. All $p\text{-values}\!\ll\!0.001$.}
  \label{tab:bias-anova}
  \small
  \begin{tabular}{lcccc}
    \toprule
    Subject & $F_\text{country}$ & $p_\text{country}$ & $F_{\text{model}\times\text{country}}$ & $p_{\text{model}\times\text{country}}$ \\
    \midrule
    CH & 17.59 & $2.5\times10^{-20}$ & 5.07 & $1.1\times10^{-11}$ \\
    CN & 14.10 & $5.0\times10^{-16}$ & 5.93 & $1.7\times10^{-14}$ \\
    LC & 17.77 & $1.5\times10^{-20}$ & 4.32 & $2.4\times10^{-9}$ \\
    MT & 12.99 & $1.1\times10^{-14}$ & 5.18 & $4.7\times10^{-12}$ \\
    \bottomrule
  \end{tabular}
\end{table}

\begin{figure}[H]
  \centering
  \includegraphics[width=1\linewidth]{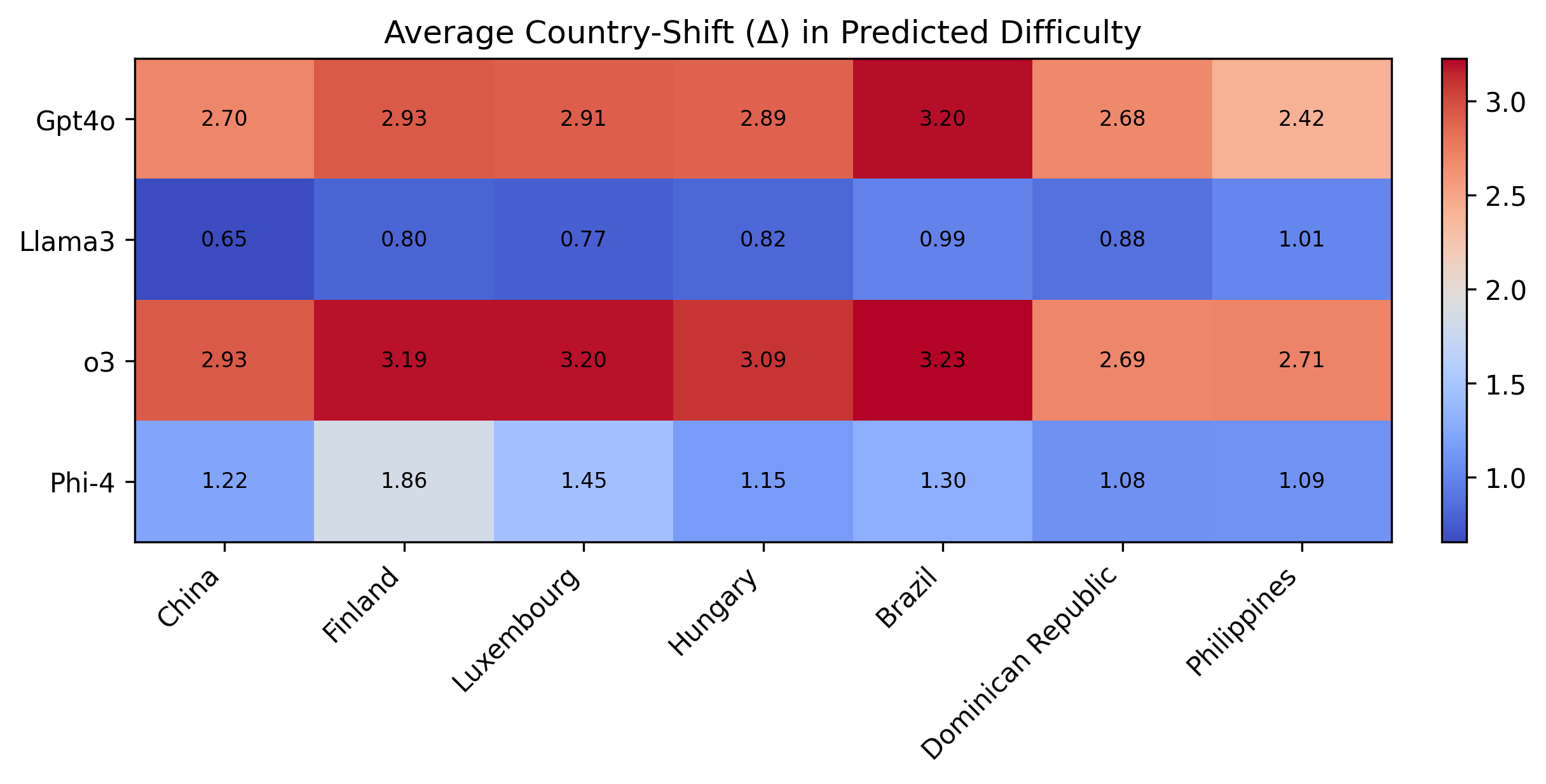}
  \caption{Overall mean $\bar{\Delta}$ (averaged over subjects) by country and model. Higher values mean stronger underestimation.}
  \label{fig:country_shift_overall}
\end{figure}

\begin{figure}[!htbp]
  \centering
  \includegraphics[width=.68\linewidth]{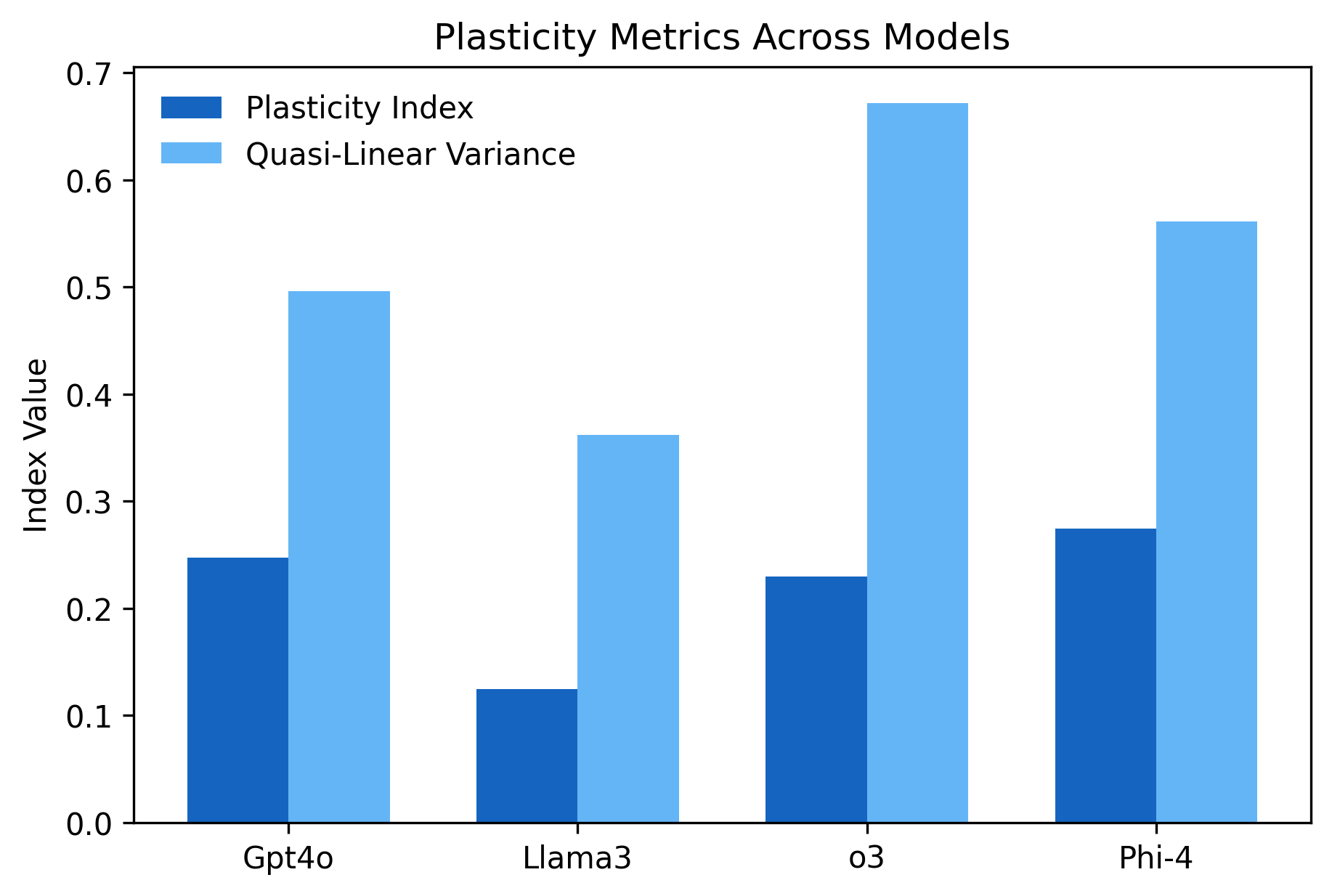}
  \caption{Plasticity metrics by model. PI is the standard deviation of country means; QLV is the average item wise variance across countries.}
  \label{fig:bias_plasticity}
\end{figure}

\noindent{\bf Responsiveness (plasticity).}
Figure~\ref{fig:bias_plasticity} reports PI and QLV. \textit{o3} shows the strongest structured response (higher QLV), \textit{Llama-3} is comparatively rigid, and \textit{Phi-4} increases PI mainly in Mathematics.

\begin{figure}[!htbp]
  \centering
  \begin{subfigure}{.50\linewidth}
    \centering\includegraphics[width=\linewidth]{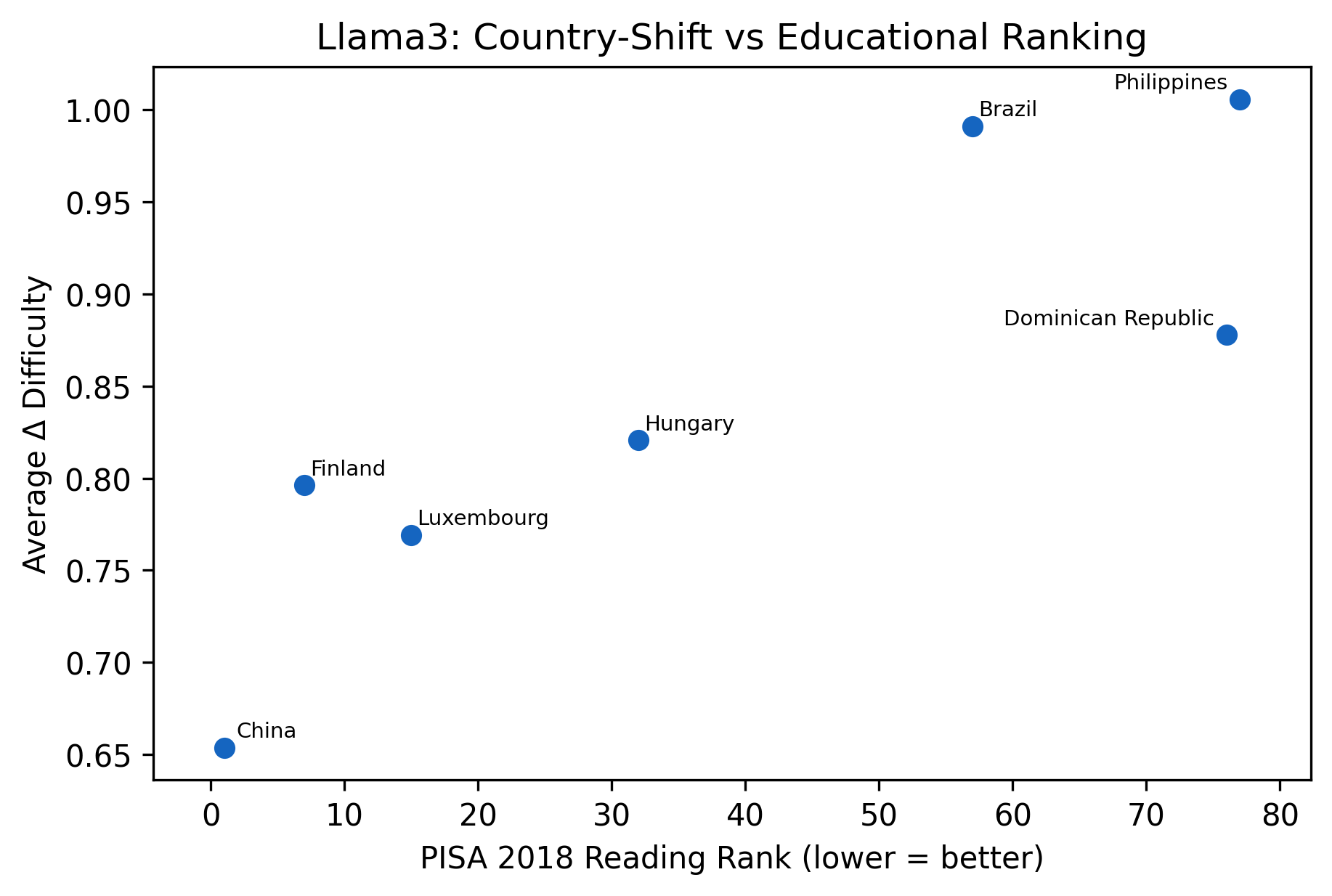}
    \caption{Llama-3}
  \end{subfigure}\hfill
  \begin{subfigure}{.50\linewidth}
    \centering\includegraphics[width=\linewidth]{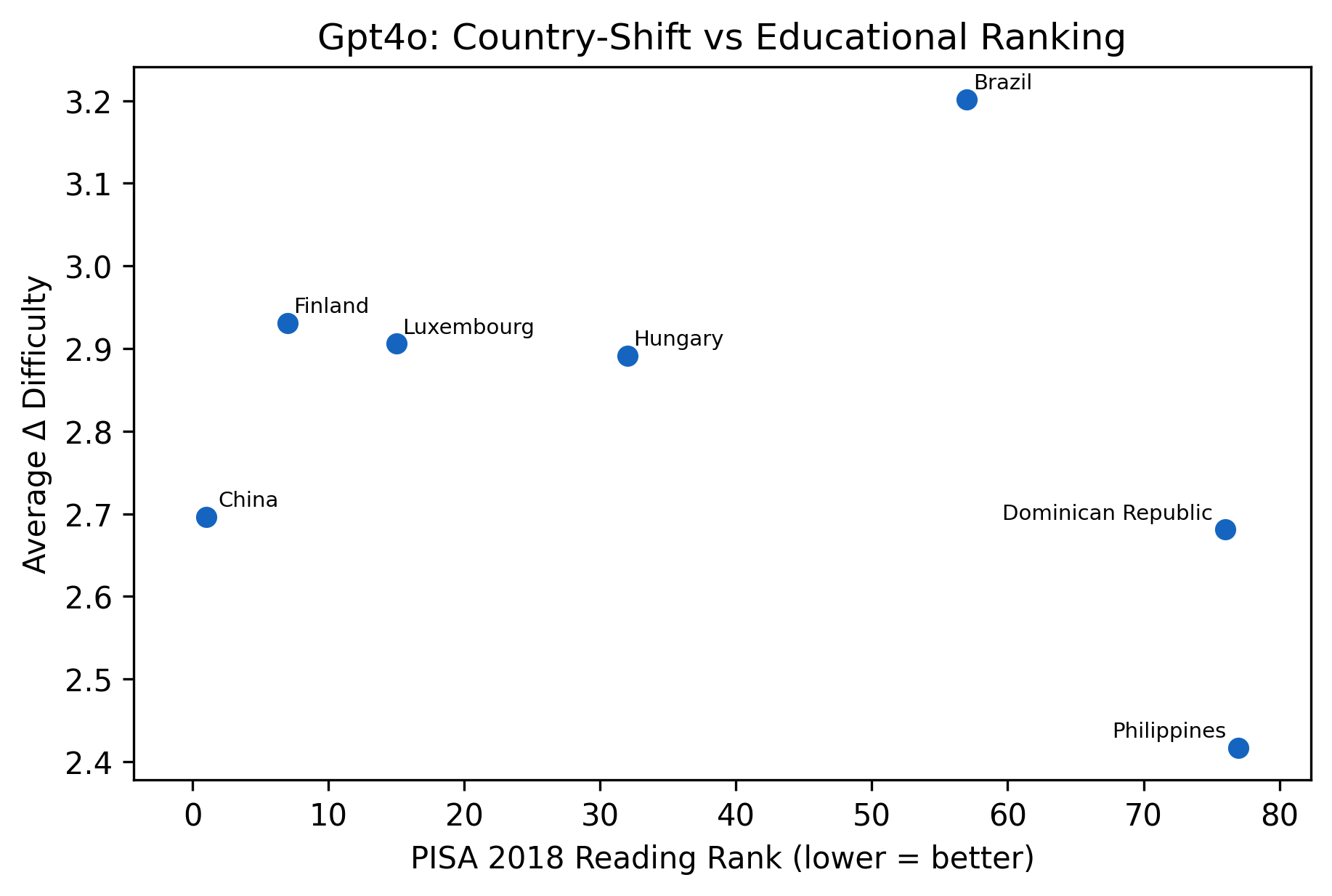}
    \caption{GPT-4o}
  \end{subfigure}\\[0.8ex]
  \begin{subfigure}{.50\linewidth}
    \centering\includegraphics[width=\linewidth]{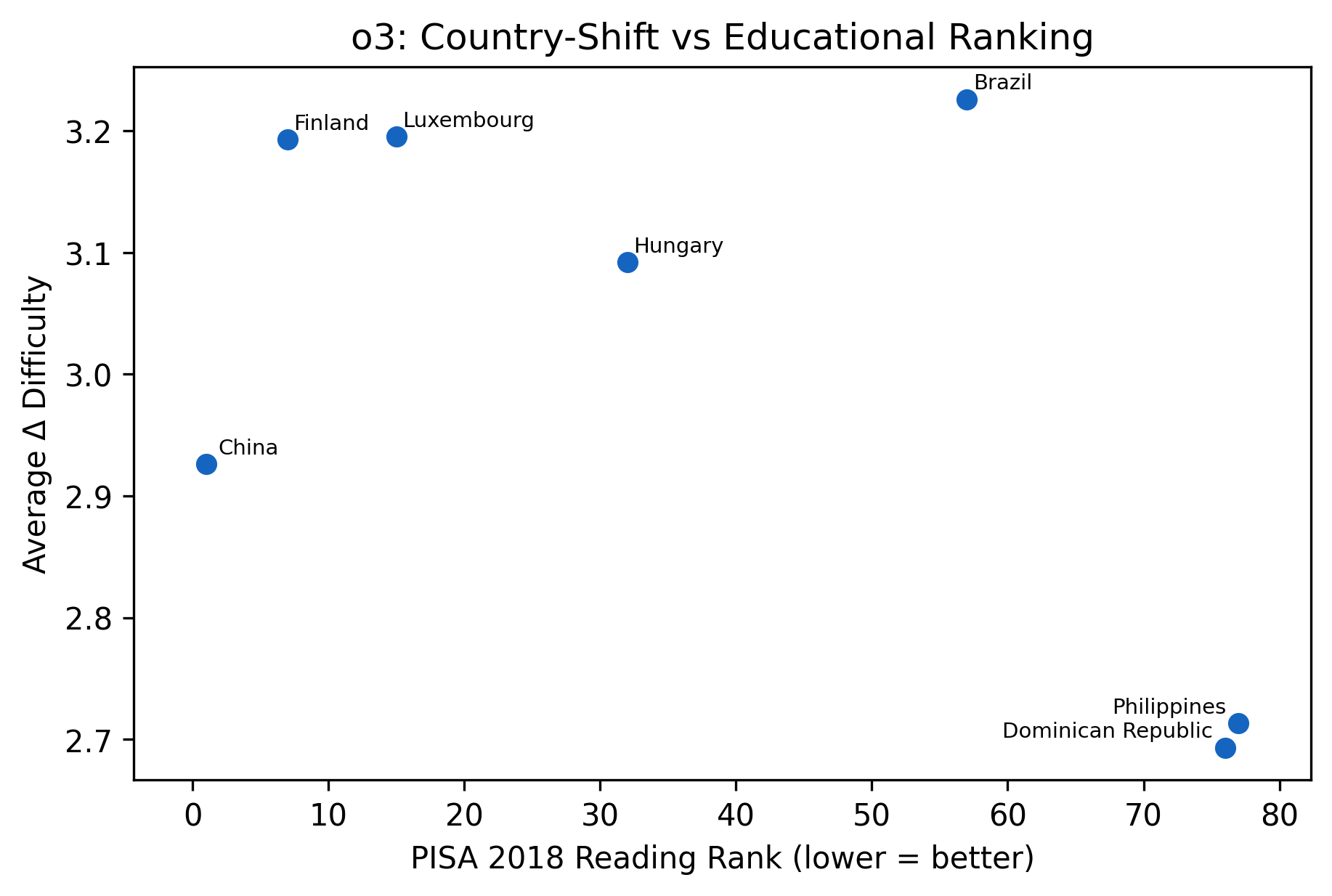}
    \caption{o3}
  \end{subfigure}\hfill
  \begin{subfigure}{.50\linewidth}
    \centering\includegraphics[width=\linewidth]{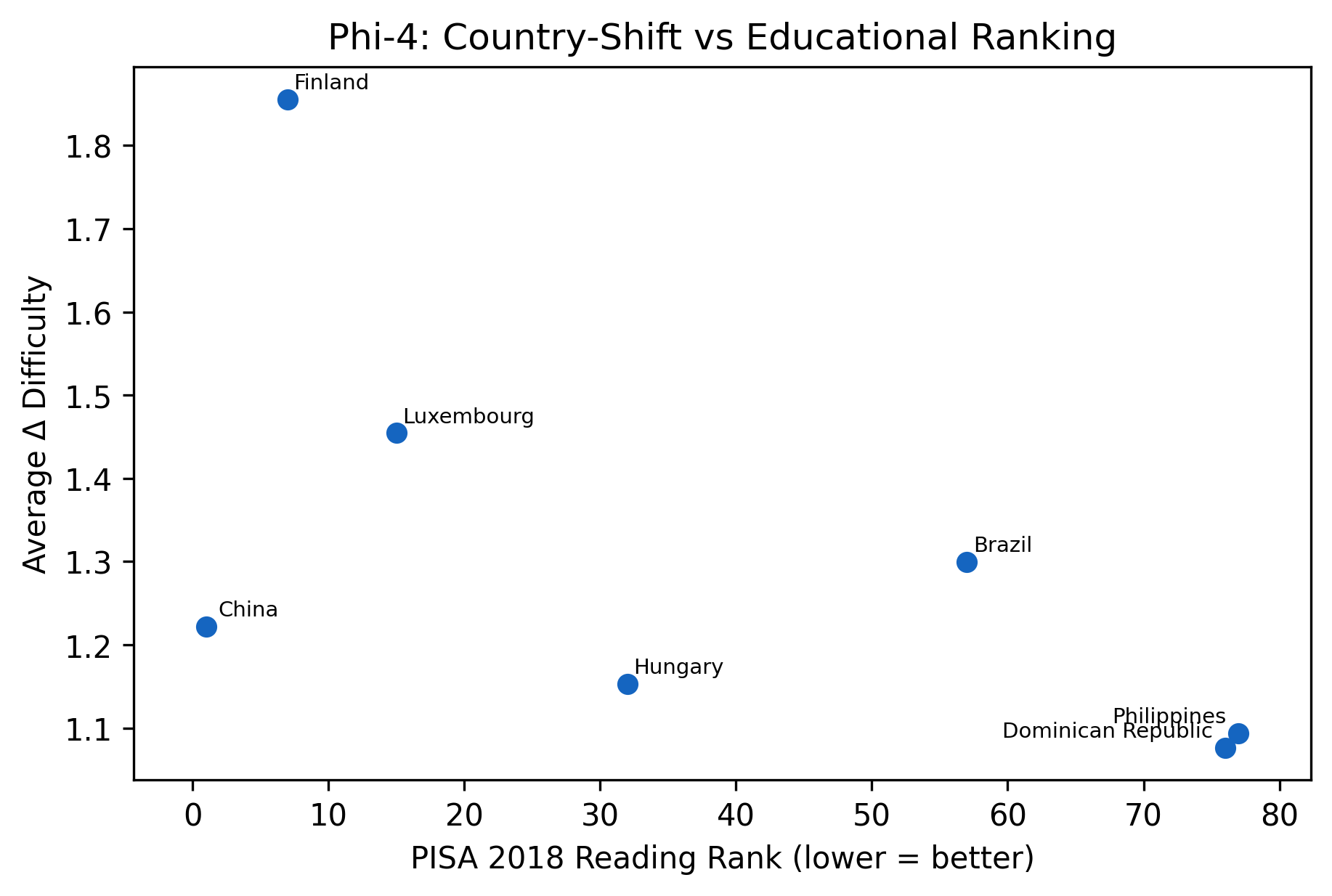}
    \caption{Phi-4}
  \end{subfigure}
  \caption{PISA Reading rank versus mean residual $\bar{\Delta}$ by country. Only Llama-3 shows a significant monotonic relationship.}
  \label{fig:bias_pisa}
\end{figure}

\smallskip
\noindent{\bf Alignment with PISA.}
We correlate each model's country means with PISA Reading rank over the seven countries. Only \textit{Llama-3} shows a significant monotonic association (Spearman $\rho=0.93$, $p=0.0025$; Kendall $\tau=0.81$, $p=0.0107$). \textit{GPT-4o} and \textit{o3} show weak negative trends; \textit{Phi-4} trends negative but is not significant (see Fig.~\ref{fig:bias_pisa}).

\paragraph{\bf Robustness Analysis.} Repeating all analyses after shift only calibration recenters residuals but leaves differences between countries largely unchanged; rank metrics are unaffected. Thus, the cue effect is not a simple location bias. Stable prompts (zero or few shot and persona few shot) show slightly smaller country spreads than reasoning heavy variants (chain or tree and point based), which sometimes amplify sensitivity, particularly in Mathematics.
Cue sensitivity concentrates in the mid difficulty band (4–6 on the 1–10 scale) and in items with visual descriptors, echoing the general modality findings.

\paragraph{\bf Guidance for Deployment.} (i) Avoid demographic or geographic tokens in production prompts. (ii) Learn a country agnostic calibration on recent labeled items and apply it before using absolute scores. (iii) Flag items with high cross country spread (high item wise variance) for review. 

\paragraph{\bf Ethics and Limitations.}
Country tokens are coarse proxies that conflate language, curriculum, and cultural context and do not represent individuals. PISA alignment is descriptive and should not be used to personalize difficulty estimates. Translation and image to text steps add noise that can interact with the cue. Reported effects should be monitored over time if deployed at scale.

A single country token produces small but consistent shifts that vary by subject and model. The largest amplification appears in Mathematics and for \textit{o3} and \textit{GPT-4o}. Only \textit{Llama-3} shows a monotonic relation with PISA ranks, and the cross country spread is small in absolute terms (about 0.35 points on the 1–10 scale). Routine diagnostics and country agnostic calibration are therefore prudent before using absolute scores.

\section{Conclusion}
\label{sec:conclusion}

\subsection{Summary of Findings}
\label{sec:conclusion-summary}
Across ten LLMs and eight prompt families on ENEM with IRT-grounded labels, we find that models recover useful rank signal but are not well calibrated in absolute terms, exhibiting systematic location/scale bias relative to IRT. Prompting choices and multimodal transcription materially shift behavior, with diagram/table items degrading more and some reasoning-heavy prompts amplifying variance. Finally, when we inject simple learner and background cues (country tokens), the resulting ``plasticity'' is limited and noisy rather than consistently aligned with real performance gaps. Therefore, difficulty estimation should be treated as a prerequisite capability that must be calibrated and audited before LLMs are used to generate or select assessment items.

This report began from a simple inversion of the current zeitgeist: while much of the educational discourse emphasizes using large language models (LLMs) to generate exam questions, we asked whether LLMs can first evaluate question difficulty---and for whom. Using Brazil's ENEM as a high-stakes, real-world testbed with IRT-grounded labels, we showed that contemporary models recover broad patterns of difficulty and usable rank signals across subjects, yet they also display systematic location/scale biases, sensitivity to prompt design and modality, and only limited, noisy adaptation to learner background cues \cite{veeramani2024llm_difficulty_rt}. These findings reinforce the storyline motivating our study: difficulty estimation is not a trivial byproduct of generation but a prerequisite capability that must be understood, calibrated, and audited before any responsible use of LLMs in assessment.

Taken together, our results support a two-tier, ``evaluation-before-generation'' pipeline. First, LLMs can serve as low-cost screeners that prioritize candidate items by predicted difficulty, provided that simple post hoc calibration corrects global biases and that evaluation emphasizes both absolute alignment with IRT metrics and the preservation of relative ordering within and across subjects \cite{karpati2023}. Second, only after items pass this evaluation gate---complete with fairness diagnostics that probe sensitivity to background cues---should they proceed to human curation and limited pilots, where psychometric fit is validated (e.g., IRT alignment, rank fidelity, and DIF checks) before operational deployment. In this framing, LLMs contribute measurable value as accelerators of item triage and quality control, but they are not yet reliable oracles of difficulty.

\subsection{Practical Guidance for Deployment}
\label{sec:conclusion-guidance}
Based on the descriptive patterns observed above (not on additional inferential tests), we recommend selecting a low-KL base model (\textit{Claude-4-Sonnet}, \textit{DeepSeek-R1}, \textit{Gemini-2.0-Flash}), scoring with a stable prompt (e.g., \texttt{persona\_few\_shot}), and selectively re-scoring mid-band items (4--6) with \texttt{point\_based} to improve sensitivity in the hard tail. When $\sigma_{\text{prompt}} > 1.3$ (\textit{GPT-4o}, \textit{Llama-3}, \textit{Mistral-Large}), single-template edits can shift labels by more than one point---apply post-hoc calibration (Section~\ref{sec:results-calibration}) and lock prompts early.

\subsection{Fairness and ``Plasticity'' Guardrails}
\label{sec:conclusion-fairness}
Our exploration of ``plasticity,'' whether LLM difficulty estimates adapt when prompts include background cues about distinct learner populations, highlights both promise and risk. We observed modest, model-specific shifts that do not consistently mirror real performance gaps. This insensitivity cautions against uncritical use of LLMs for adaptive testing or equity-sensitive personalization: insufficient plasticity risks a one-size-fits-all notion of difficulty, while spurious plasticity risks encoding bias and amplifying inequalities \cite{li2023fairnessllms}. Therefore, any deployment must include explicit fairness diagnostics and guardrails, with clear documentation of when and how contextual information is allowed to influence difficulty estimates \cite{mitchell2019model}.

\subsection{Limitations and Outlook}
\label{sec:conclusion-limitations-outlook}
Like any technical investigation, our study has limitations. We focus on released ENEM items from a fixed time window (2017--2022); accordingly, some effects may be year- and format-dependent, and generalization to other languages, curricula, or item formats remains an open question. In particular, the magnitude of global location/scale bias (and thus the benefits of post hoc calibration) may drift with changes in test design or cohort composition across years, and the modality penalty may vary with the prevalence and style of diagrams and tables in different exam forms. Most models received text-only representations of visuals, which likely understates the potential of native vision--language systems that preserve layout and spatial structure. Finally, our plasticity probes used ethically minimal background cues; richer, better-grounded contextualization may elicit different behavior, but it also raises nontrivial privacy, consent, and bias concerns that must be treated as first-class design constraints \cite{regan2019ethical}.

These limitations motivate a concrete research agenda. Psychometric alignment at scale, combining LLM signals with small pilot samples to learn stable mappings to IRT metrics across cohorts and years, remains a priority \cite{settles2020machine}. Vision--language fidelity should be evaluated with models that preserve spatial structure and layout, especially for diagram-heavy questions \cite{lu2022learn}. Fairness work must move beyond cue-based prompts to counterfactual and DIF-style evaluations, quantifying when adaptation is legitimate and when it is harmful \cite{li2023fairnessllms}. Finally, closed-loop pipelines that integrate retrieval of curricular references, controlled generation, and a pre-registered ``evaluation gate'' can turn LLMs into auditable components of trustworthy assessment systems rather than opaque end-to-end replacements \cite{raji2020closing}.

In summary, LLMs already provide actionable signal for screening the difficulty of exam questions, but they are not yet dependable, stand-alone estimators across subjects, modalities, and student populations. The path to responsible use runs through calibration, rank-aware evaluation, preservation of visual structure, and explicit fairness diagnostics. Treating difficulty estimation and equity auditing as foundational---not afterthoughts---will be essential to any future pipeline that hopes to leverage LLMs for the safe and equitable generation of new assessment content.



\bibliographystyle{ieeetr}
\bibliography{references}

\newpage
\appendix
\section{Prompts}
\label{appendix:prompts}
\begin{table}[htbp]
    \begin{center} 
        \caption{Zero-Shot Prompt}
        \label{table:zero-shot}
        \begin{tabular}{p{\linewidth}}
            \hline
            \textbf{System Content} \\
            Classify the difficulty of the following exam question on a scale of 1 to 10, where 1 is very easy and 10 is very hard: \\
            \hline
            \textbf{Complement} \\
            Answer only with an integer representing the difficulty. \\
            \hline
        \end{tabular}
    \end{center}
\end{table}

\begin{table}[htbp]
    \begin{center} 
        \caption{Enhanced Zero-Shot Prompt}
        \label{table:enhanced-zero-shot}
        \begin{tabular}{p{\linewidth}}
            \hline
            \textbf{System Content} \\
            Classify the difficulty of the following exam question from the ENEM, the national exam in Brazil, and consider the difficulty for the average student who will take the exam. On a scale of 1 to 10, where 1 is a very easy question (almost all the students answered correctly) and 10 is a very hard question (almost all the students did not answer correctly): \\
            \hline
            \textbf{Complement} \\
            Answer only with an integer representing the difficulty. \\
            \hline
        \end{tabular}
    \end{center}
\end{table}

\begin{table}[htbp]
    \begin{center} 
        \caption{Persona + Few-Shot Prompt}
        \label{table:persona+few-shot}
        \begin{tabular}{p{\linewidth}}
            \hline
            \textbf{System Content} \\
            As an education specialist, your task is to help me create suitable questions for Brazilian high school students in their last year. The questions should be catered to test their level of high school knowledge.\\
            The ENEM exam is a standardized Brazilian national exam, which evaluates high
            school students in Brazil. The exam has been used as an admission to universities.  \\

            \textbf{Question Difficulty Examples} \\ 
            Here are two examples of past ENEM exams: \\

            \hspace{10pt} \textit{Example 1: \{A hard question and its corresponding answer options.\}} \\ 
            \hspace{10pt} This question Difficulty is \{9\}. \\

            \vspace{2pt} 

            \hspace{10pt} \textit{Example 2: \{An easy question and its corresponding answer options.\}} \\ 
            \hspace{10pt} This question Difficulty is \{2\}. \\

            Given these examples, evaluate the difficulty of the following question. \\
            \hline
            \textbf{Complement} \\
            Answer only with an integer representing the difficulty. \\
            \hline
        \end{tabular}
    \end{center}
\end{table}

\begin{table}[htbp]
    \begin{center} 
        \caption{Few-Shot Prompt}
        \label{table:few-shot}
        \begin{tabular}{p{\linewidth}}
            \hline
            \textbf{System Content} \\
            Classify the difficulty of the following exam question on a scale of 1 to 10, where 1 is very easy and 10 is very hard: \\

            \textbf{Question Difficulty Examples} \\ 
            Here are two examples of past exams: \\

            \hspace{10pt} \textit{Example 1: \{A hard question and its corresponding answer options.\}} \\ 
            \hspace{10pt} This question Difficulty is \{9\}. \\

            \vspace{2pt} 

            \hspace{10pt} \textit{Example 2: \{An easy question and its corresponding answer options.\}} \\ 
            \hspace{10pt} This question Difficulty is \{2\}. \\

            Given these examples, evaluate the difficulty of the following question. \\
            \hline
            \textbf{Complement} \\
            Answer only with an integer representing the difficulty. \\
            \hline
        \end{tabular}
    \end{center}
\end{table}

\begin{table}[htbp]
    \begin{center} 
        \caption{Chain of Thought Prompt}
        \label{table:chainofhtought}
        \begin{tabular}{p{\linewidth}}
            \hline
            \textbf{System Content} \\
            Classify the difficulty of the following exam question on a scale of 1 to 10, where 1 is very easy and 10 is very hard:  \\

            \textbf{Question Difficulty Examples} \\ 
            Here are two examples of past ENEM exams: \\

            \hspace{10pt} \textit{Example 1: \{A hard question and its corresponding answer options.\}} \\ 
            \hspace{10pt} This question Difficulty is \{9\}. \\

            \vspace{2pt} 

            \hspace{10pt} \textit{Example 2: \{An easy question and its corresponding answer options.\}} \\ 
            \hspace{10pt} This question Difficulty is \{2\}. \\

            Given these examples, evaluate the difficulty of the following question. \\
            \hline
            \textbf{Complement} \\
            Step 1: Try to answer the question.\\
            Step 2: Explain where the difficulty of the question lies. \\
            Step 3: Compare with the two examples. \\
            Step 4: Rate the difficulty of the question on a scale of 1 (very easy) to 10 (very hard). \\
            Step 5: Rate your confidence in percentage for your answer.\\
            Step 6: Format your answer as follows:\\

            \hspace{10pt}\textit{Answer: $<$answer$>$}
            
            \hspace{10pt}\textit{Explain: $<$explain$>$}
            
            \hspace{10pt}\textit{Compare with the two examples: $<$comparison$>$}
            
            \hspace{10pt}\textit{Difficulty: $<$difficulty$>$}
            
            \hspace{10pt}\textit{Confidence: $<$confidence$>$} \\
            \hline
        \end{tabular}
    \end{center}
\end{table}

\begin{table}[htbp]
    \begin{center} 
        \caption{Point based Prompt}
        \label{table:pointbased}
        \begin{tabular}{p{\linewidth}}
            \hline
            \textbf{System Content} \\
            You are an expert ENEM exam creator and grader. Your task is to assess the difficulty of ENEM exam questions on a scale of 1-10, where: \\
            \vspace{1pt}
            1 = Extremely easy (nearly all students answer correctly)\\
            10 = Extremely difficult (very few students answer correctly)\\
            \vspace{1pt}
            
            Important: Use the full range of scores from 1-10. Avoid overusing middle scores.\\
            Here is the question to evaluate: \\
            \hline
            \textbf{Complement} \\
            Main skill tested: [Brief description]\\
            Estimated percentage correct: [Your estimate]\\
            Initial difficulty score: [1-10 based on percentage]\\
            \vspace{1pt}
            Adjusting factors:\\
            - Language complexity: [Simple/Moderate/Complex] [+0/+1/+2]\\
            - Steps required: [Few/Several/Many] [+0/+1/+2]\\
            - Concept abstractness: [Concrete/Mixed/Abstract] [+0/+1/+2]\\
            - Question uniqueness: [Common/Uncommon/Rare] [+0/+1/+2]\\
            \vspace{1pt}
            Final difficulty score: [1-10, adjusted from initial score]\\
            Explanation: [Brief justification of your final score]\\
            \hline
        \end{tabular}
    \end{center}
\end{table}

\begin{table}[htbp]
    \begin{center} 
        \caption{Tree of Thoughts Prompt}
        \label{table:treeofthoughts}
        \begin{tabular}{p{\linewidth}}
            \hline
            \textbf{System Content} \\
            Classify the difficulty of the following exam question on a scale of 1 to 10, where 1 is very easy and 10 is very hard.  \\

            \textbf{Question Difficulty Examples} \\ 
            Here are two examples of past ENEM exams: \\

            \hspace{10pt} \textit{Example 1: \{A hard question and its corresponding answer options.\}} \\ 
            \hspace{10pt} This question Difficulty is \{9\}. \\

            \vspace{2pt} 

            \hspace{10pt} \textit{Example 2: \{An easy question and its corresponding answer options.\}} \\ 
            \hspace{10pt} This question Difficulty is \{2\}. \\
            
            \vspace{2pt}
            
            Now imagine three different ENEM exam creators and graders are evaluating the difficulty of ENEM exam questions on a scale of 1 to 10. All experts will write down step 1 of their thinking. Then, share it with the group. Then, all experts will go on to the next step. If any expert realizes they are wrong at any point, then they leave. \\
            \vspace{1pt}
            What is the difficulty of the question? \\
            \vspace{1pt}
            Important: Use the full range of scores from 1-10. Avoid overusing middle scores. \\
            \hline
            \textbf{Complement} \\
            Step 1: Try to answer the question.\\
            Step 2: Explain where the difficulty of the question lies. \\
            Step 3: Compare with the two examples. \\
            Step 4: Rate the difficulty of the question on a scale of 1 (very easy) to 10 (very hard). \\
            Step 5: Rate your confidence in percentage for your answer.\\
            Step 6: Format your answer as follows:\\
            \vspace{1pt}
            \hspace{10pt}\textit{Answer: $<$answer$>$}
            
            \hspace{10pt}\textit{Explain: $<$explain$>$}
            
            \hspace{10pt}\textit{Compare with the two examples: $<$comparison$>$}
            
            \hspace{10pt}\textit{Difficulty: $<$difficulty$>$}
            
            \hspace{10pt}\textit{Confidence: $<$confidence$>$} \\
            \hline
        \end{tabular}
    \end{center}
\end{table}

\begin{table}[htbp]
    \begin{center} 
        \caption{Synthetic Prompt}
        \label{table:playground}
        \begin{tabular}{p{\linewidth}}
            \hline
            \textbf{System Content} \\
            Evaluate the difficulty level of questions from the ENEM (Exame Nacional do Ensino M\'edio), the national exam in Brazil, on a scale from 1 (easy) to 10 (hard). \\
            Consider the complexity, depth of understanding required, and the skills needed to answer each question effectively. Your evaluation should reflect an objective and balanced assessment based on the described criteria. \\

            \textbf{Question Difficulty Examples} \\
            \# Examples \\
            \vspace{1pt}
            \textit{**Example 1**:} \\
            \vspace{1pt}
            \textit{**Input**: A hard question and its corresponding answer options.} \\
            \vspace{1pt}
            \textit{**Output**:} \\
            \texttt{\{ \newline}
            \hspace{10pt} \texttt{"difficulty\_level\: 9,} \newline
            \hspace{10pt} \texttt{"rationale": "The rationale behind why the question is considered difficult"} \newline
            \texttt{\}} \\

            \vspace{2pt} 

            \textit{**Example 2**:} \\
            \vspace{1pt}
            \textit{**Input**: An easy question and its corresponding answer options.} \\
            \textit{**Output**:} \\
            \texttt{\{ \newline}
            \hspace{10pt} \texttt{"difficulty\_level\: 2,} \newline
            \hspace{10pt} \texttt{"rationale": "The rationale behind why the question is considered easy"} \newline
            \texttt{\}} \\
            \hline
            \textbf{Complement} \\
            \# Steps \\
            1. **Understand Question Context**: Analyze the subject matter and any potential subtopics covered in the question. \\
            2. **Evaluate Complexity**: Consider the logical reasoning, calculations, or critical thinking needed to solve the question. \\
            3. **Assess Required Skills**: Identify the type of skills (e.g., analytical, mathematical, language) needed and their proficiency level. \\
            4. **Compare Against Standards**: Consider how the question compares to typical educational standards for high school graduates. \\
            5. **Assign Difficulty Level**: Based on your analysis, assign a difficulty score from 1 to 10, with 1 being the easiest and 10 the most difficult. \\
            \vspace{1pt}
            \# Output Format \\
            Provide the evaluation as a JSON object with the following structure: \\
            \texttt{\{ \newline}
            \hspace{10pt} \texttt{"difficulty\_level\: [1-10],} \newline
            \hspace{10pt} \texttt{"rationale": "Provide a brief explanation of the reasoning behind the difficulty level assigned."} \newline
            \texttt{\}} \\
            \vspace{1pt}
            \#Notes \\
            - Be mindful of different subject areas and cognitive requirements when evaluating questions. \\
            - Consider cultural and contextual differences relevant to the Brazilian education system. \\
            \hline
        \end{tabular}
    \end{center}
\end{table}

\clearpage
\section{Complete model-prompt results}
\label{app:overall-table}

Each table lists the eight prompts for a single LLM together with every
metric used in this thesis.  
Error metrics (RMSE, MAE) are minimised; accuracy / correlation metrics
(Within-1 accuracy, Exact-Match, Spearman $\rho$) are
maximised.  The best value per column is highlighted in \textbf{bold}.

\paragraph{Prompt abbreviations.}
CoT = \texttt{chain\_of\_thought}, EZ = \texttt{enhanced\_zero\_shot},
FS = \texttt{few\_shot}, PFS = \texttt{persona\_few\_shot},
PB = \texttt{point\_based}, ToT = \texttt{tree\_of\_thought},
ZS = \texttt{zero\_shot}.

\section{Claude-4-Sonnet}

\begin{tabular}{lccccc}
\toprule
Prompt & RMSE & MAE & W1A & EM & $\rho$ \\
\midrule
CoT & 2.50 & 1.84 & 0.55 & 0.22 & 0.05 \\
EZ  & \textbf{1.83} & \textbf{1.44} & 0.58 & 0.20 & \textbf{0.33} \\
FS  & 2.05 & 1.45 & \textbf{0.63} & \textbf{0.29} & 0.22 \\
PFS & 2.08 & 1.52 & 0.60 & 0.26 & 0.25 \\
PG  & --   & --   & --   & --   & -- \\
PB  & --   & --   & --   & --   & -- \\
ToT & --   & --   & --   & --   & -- \\
ZS  & --   & --   & --   & --   & -- \\
\bottomrule
\end{tabular}

\section{DeepSeek-R1}

\begin{tabular}{lccccc}
\toprule
Prompt & RMSE & MAE & W1A & EM & $\rho$ \\
\midrule
CoT & 2.17 & 1.54 & 0.62 & 0.27 & 0.13 \\
EZ  & 1.82 & 1.40 & 0.60 & 0.23 & \textbf{0.35} \\
FS  & 2.09 & 1.52 & 0.61 & 0.25 & 0.12 \\
PFS & \textbf{1.78} & \textbf{1.31} & \textbf{0.66} & \textbf{0.27} & 0.15 \\
PG  & 1.90 & 1.41 & 0.63 & 0.24 & 0.20 \\
PB  & 1.98 & 1.58 & 0.52 & 0.19 & 0.20 \\
ToT & 1.97 & 1.41 & 0.64 & 0.27 & 0.13 \\
ZS  & 2.04 & 1.60 & 0.55 & 0.18 & 0.21 \\
\bottomrule
\end{tabular}

\section{Gemini-2.0-Flash}

\begin{tabular}{lccccc}
\toprule
Prompt & RMSE & MAE & W1A & EM & $\rho$ \\
\midrule
CoT & 1.98 & 1.48 & 0.60 & 0.24 & 0.15 \\
EZ  & 2.25 & 1.86 & 0.43 & 0.13 & \textbf{0.33} \\
FS  & 1.81 & 1.32 & 0.66 & \textbf{0.27} & 0.17 \\
PFS & 1.81 & 1.34 & 0.66 & 0.25 & 0.15 \\
PG  & 2.33 & 1.78 & 0.51 & 0.20 & 0.23 \\
PB  & 2.19 & 1.84 & 0.41 & 0.12 & 0.32 \\
ToT & 2.06 & 1.55 & 0.59 & 0.22 & 0.07 \\
ZS  & \textbf{1.70} & \textbf{1.29} & \textbf{0.66} & 0.24 & 0.20 \\
\bottomrule
\end{tabular}

\section{Gemini-2.5-Flash}

\begin{tabular}{lccccc}
\toprule
Prompt & RMSE & MAE & W1A & EM & $\rho$ \\
\midrule
CoT & 2.54 & 1.90 & 0.52 & 0.19 & 0.17 \\
EZ  & 2.43 & 2.00 & 0.40 & 0.13 & \textbf{0.41} \\
FS  & 2.33 & 1.77 & 0.52 & 0.21 & 0.14 \\
PFS & \textbf{1.93} & \textbf{1.44} & \textbf{0.62} & \textbf{0.24} & 0.18 \\
PG  & 2.20 & 1.72 & 0.51 & 0.19 & 0.31 \\
PB  & 1.97 & 1.49 & 0.59 & 0.22 & 0.37 \\
ToT & 2.65 & 2.02 & 0.47 & 0.19 & 0.26 \\
ZS  & 2.12 & 1.70 & 0.50 & 0.17 & 0.31 \\
\bottomrule
\end{tabular}

\section{GPT-4o}

\begin{tabular}{lccccc}
\toprule
Prompt & RMSE & MAE & W1A & EM & $\rho$ \\
\midrule
CoT & 3.74 & 3.33 & 0.16 & 0.05 & 0.15 \\
EZ  & 3.50 & 3.17 & 0.13 & 0.03 & \textbf{0.33} \\
FS  & 3.64 & 3.17 & 0.20 & 0.07 & 0.17 \\
PFS & 3.53 & 3.07 & 0.21 & 0.07 & 0.20 \\
PG  & 3.74 & 3.38 & 0.13 & 0.04 & 0.22 \\
PB  & \textbf{2.25} & \textbf{1.78} & \textbf{0.49} & \textbf{0.18} & 0.31 \\
ToT & 3.76 & 3.38 & 0.13 & 0.03 & 0.20 \\
ZS  & 3.83 & 3.51 & 0.11 & 0.02 & 0.24 \\
\bottomrule
\end{tabular}

\section{Llama-3}

\begin{tabular}{lccccc}
\toprule
Prompt & RMSE & MAE & W1A & EM & $\rho$ \\
\midrule
CoT & 2.92 & 2.35 & 0.37 & 0.15 & -0.08 \\
EZ  & 1.60 & 1.25 & 0.65 & 0.24 & -0.01 \\
FS  & \textbf{1.40} & \textbf{1.05} & \textbf{0.73} & \textbf{0.30} & \textbf{0.10} \\
PFS & 2.12 & 1.55 & 0.59 & 0.23 & -0.21 \\
PG  & 3.15 & 2.59 & 0.34 & 0.14 & -0.15 \\
PB  & 1.73 & 1.32 & 0.64 & 0.24 & -0.03 \\
ToT & 2.57 & 2.00 & 0.47 & 0.17 & -0.08 \\
ZS  & 1.66 & 1.31 & 0.63 & 0.23 & -0.02 \\
\bottomrule
\end{tabular}

\section{Mistral-Large}

\begin{tabular}{lccccc}
\toprule
Prompt & RMSE & MAE & W1A & EM & $\rho$ \\
\midrule
CoT & 2.09 & \textbf{1.47} & \textbf{0.64} & \textbf{0.26} & 0.03 \\
EZ  & \textbf{2.00} & 1.60 & 0.54 & 0.17 & \textbf{0.34} \\
FS  & 2.39 & 1.81 & 0.52 & 0.22 & 0.14 \\
PFS & 2.04 & 1.50 & 0.60 & 0.25 & 0.15 \\
PG  & 2.29 & 1.80 & 0.50 & 0.16 & 0.20 \\
PB  & 2.49 & 2.15 & 0.31 & 0.10 & 0.23 \\
ToT & 2.59 & 1.94 & 0.52 & 0.21 & 0.02 \\
ZS  & 2.05 & 1.61 & 0.54 & 0.18 & 0.26 \\
\bottomrule
\end{tabular}

\section{o3}

\begin{tabular}{lccccc}
\toprule
Prompt & RMSE & MAE & W1A & EM & $\rho$ \\
\midrule
CoT & 3.74 & 3.33 & 0.16 & 0.05 & 0.15 \\
EZ  & 3.50 & 3.17 & 0.13 & 0.03 & \textbf{0.33} \\
FS  & 3.64 & 3.17 & 0.20 & 0.07 & 0.17 \\
PFS & 3.53 & 3.07 & 0.21 & 0.07 & 0.20 \\
PG  & 3.74 & 3.38 & 0.13 & 0.04 & 0.22 \\
PB  & \textbf{2.25} & \textbf{1.78} & \textbf{0.49} & \textbf{0.18} & 0.31 \\
ToT & 3.76 & 3.38 & 0.13 & 0.03 & 0.20 \\
ZS  & 3.83 & 3.51 & 0.11 & 0.02 & 0.24 \\
\bottomrule
\end{tabular}

\section{Phi-4}

\begin{tabular}{lccccc}
\toprule
Prompt & RMSE & MAE & W1A & EM & $\rho$ \\
\midrule
CoT & 2.01 & 1.41 & 0.64 & 0.29 & 0.20 \\
EZ  & 2.68 & 2.23 & 0.37 & 0.11 & 0.16 \\
FS  & 1.63 & 1.17 & 0.71 & 0.29 & \textbf{0.34} \\
PFS & \textbf{1.47} & \textbf{1.06} & \textbf{0.74} & \textbf{0.32} & 0.31 \\
PG  & 1.78 & 1.27 & 0.67 & 0.29 & 0.32 \\
PB  & 1.97 & 1.55 & 0.57 & 0.18 & 0.09 \\
ToT & 2.48 & 1.91 & 0.48 & 0.19 & 0.16 \\
ZS  & 2.68 & 2.14 & 0.42 & 0.15 & 0.12 \\
\bottomrule
\end{tabular}

\section{Qwen-3}

\begin{tabular}{lccccc}
\toprule
Prompt & RMSE & MAE & W1A & EM & $\rho$ \\
\midrule
CoT & 1.91 & 1.49 & 0.57 & 0.21 & 0.07 \\
EZ  & 2.30 & 1.89 & 0.44 & 0.14 & 0.15 \\
FS  & 1.91 & 1.52 & 0.57 & 0.18 & 0.08 \\
PFS & 1.91 & 1.51 & 0.56 & 0.20 & 0.11 \\
PG  & 2.15 & 1.73 & 0.47 & 0.17 & \textbf{0.18} \\
PB  & 2.34 & 1.86 & 0.45 & 0.18 & 0.12 \\
ToT & \textbf{1.84} & \textbf{1.38} & \textbf{0.64} & \textbf{0.23} & 0.07 \\
ZS  & 2.00 & 1.57 & 0.55 & 0.19 & 0.12 \\
\bottomrule
\end{tabular}

\section{Mean-predictor baseline}

\begin{tabular}{lccccc}
\toprule
Prompt & RMSE & MAE & W1A & EM & $\rho$ \\
\midrule
-- & \textbf{1.19} & \textbf{0.88} & \textbf{0.79} & \textbf{0.35} & \textbf{0.00} \\
\bottomrule
\end{tabular}

\end{document}